\documentclass[reprint,amsmath,amssymb,aps]{revtex4-1}
\usepackage{graphicx}
\usepackage{dcolumn}
\usepackage{bm}
\usepackage{here}

\def\a{\alpha}
\def\b{\beta}

\def\d{\delta}
\def\D{\Delta}

\def\br{\boldsymbol{r}}

\def\be{\boldsymbol{e}}

\def\br{\boldsymbol{r}}

\def\bkappa{{\boldsymbol{\kappa}}}

\def\rmref{{\rm ref}}

\def\ro{{\text{o}}}

\def\rc{{\text{c}}}
\def\rd{{\rm{d}}}

\def\re{{\text{e}}}

\def\ro{{\rm{o}}} 
\def\rp{{\rm{p}}}

\def\rR{{\rm{R}}}
\def\rs{{\rm{s}}}

\def\rL{{\text{L}}}

\def\rp{{\text{p}}}

\def\rR{{\text{R}}}
\def\rs{{\text{s}}}

\def\be{\boldsymbol{e}}

\def\Br{{\sf Br}}

\def\Pe{{\sf Pe}}
\def\Re{{\sf Re}}
\def\Wi{{\sf Wi}}

\def\Dt{{\Delta t}}

\def\tt{{\tilde{t}}}

\def\Re{{\sf Re}}

\begin{document}
\title{%
Multiscale simulation of a polymer melt flow \\
between two coaxial cylinders under nonisothermal conditions
}
\author{%
Yuji Hamada
} 
\affiliation{%
Graduate School of Engineering, Kyoto University\\
Katsura Campus, Nishikyo-ku, Kyoto 615-8510, JAPAN
}
\author{%
Takeshi Sato
} 
\affiliation{%
Graduate School of Engineering, Kyoto University\\
Katsura Campus, Nishikyo-ku, Kyoto 615-8510, JAPAN
}
\author{%
Takashi 
Taniguchi
} 
\email{taniguch@cheme.kyoto-u.ac.jp}
\affiliation{%
Graduate School of Engineering, Kyoto University\\
Katsura Campus, Nishikyo-ku, Kyoto 615-8510, JAPAN
}
\affiliation{%
Department of Physics,
Tohoku University, Sendai, Miyagi
980-8578, Japan
}
\begin{abstract}
We successfully extend a multiscale simulation (MSS) method
to nonisothermal well-entangled polymer melt flows
between two coaxial cylinders.
In the multiscale simulation, the macroscopic flow system
is connected to a number of microscopic systems
through the velocity gradient tensor, stress tensor and temperature.
At the macroscopic level, 
in addition to the momentum balance equation,
we consider the energy balance equation, where heat generation
plays an important role not only in the temperature distribution 
but also in the flow profile. 
At the microscopic level,
a dual slip-link model is employed for well-entangled polymers.
To incorporate the temperature effect into the microscopic systems,
we used the time-temperature superposition rule for the slip-link model, 
in which the temperature dependence of the parameters is not known;
on the other hand, the way to take into account the temperature effect
in the macroscopic equations has been well established.
We find that the extended multiscale simulation method
is quite effective in revealing the relation between nonisothermal
polymeric flows
for both steady and transient cases
and the microscopic states of polymer chains 
expressed by primitive paths and slip-links.
It is also found that the temperature-dependent reptation-time-based Weissenberg number
is a suitable measure for understanding the extent of the polymer chain deformation
in the range of the shear rate used in this study.
\end{abstract}
                           \maketitle

\section{\label{sec:Intro}Introduction}

Polymeric material has found increasingly more applications, such
as automobile, fiber, medical and aerospace materials, 
where it is needed to make polymeric products with high functionalities.
To fabricate polymer products, we usually have to address
entangled polymer melts.
In general, it is difficult to accurately predict processing flow
properties because the transport of mass, 
momentum and energy occur simultaneously in complex flow channel
geometries.
Moreover, the flow behavior of entangled polymer melts themselves is
complex because the macroscopic flow behavior is highly correlated
with the microscopic states of polymer chains, 
such as chain orientations and entanglements\cite{Masubuchi2014,Sato2020}. 

In this paper,
we specifically focus on nonisothermal flows of entangled polymer melts
that are frequently seen in the polymer industry.
Conventionally, macroscopic approaches are used to simulate
the processing flows of entangled polymer melts with temperature changes.
In the majority of macroscopic approaches, macroscopic balance equations
including the energy balance equation are coupled with
a phenomenological constitutive equation with a temperature-dependent
relaxation time.
One of the phenomenological constitutive equations
frequently employed in the polymer industry is
the generalized Newtonian fluid constitutive equation\cite{Bird1987}. 
This type of constitutive equation, e.g., the Cross model\cite{Cross1965}, 
can reproduce shear-rate-dependence of the viscosity of polymer melts. 
Furthermore, to address the effect of temperature changes, 
the generalized Newtonian fluid constitutive equation,
including the Cross model, was extended
using the Williams-Landel-Ferry (WLF) equation\cite{Ferry1980} or the Arrhenius law, 
which is the so-called Cross-WLF model or Cross-Arrhenius model, respectively. 
In the Cross-WLF model (or the Cross-Arrhenius model), 
the relaxation time and zero shear viscosity of the polymer melt
can be considered to depend on the temperature, 
which is described by the WLF equation (or the Arrhenius equation).
Using the generalized Newtonian fluid constitutive equation, 
processing flow problems with temperature changes,
such as the filling process\cite{Chiang1991, Otmani2011}, 
flows in an extruder\cite{Khalifeh2005}, 
and flows in a sudden expansion channel\cite{Zdanski2009}, were
investigated.
Although the generalized Newtonian constitutive equations
with temperature changes can reproduce the shear-rate-
and temperature-dependence of the viscosity of polymeric liquids, 
several properties of polymeric liquids, 
such as normal stress effects, are not taken into account. 
More realistically, 
a viscoelastic constitutive equation
with a temperature-dependent relaxation time replaces
the generalized Newtonian constitutive equations. 
Viscoelastic constitutive equations can successfully address normal
stress effects or time-dependent effects.
By means of these phenomenological constitutive equations, 
polymer processing flows with temperature changes, 
such as flows in a polymer melt spinning process\cite{Doufas2000}, 
around a cylinder\cite{Peters1997}, 
and in an abrupt contraction channel\cite{Kunisch2000, Wachs2000},
were extensively investigated. 
However, there are several problems
when employing phenomenological constitutive equations.
One of these problems is that phenomenological constitutive equations
prohibit one from directly obtaining microscopic insights of polymer chains. 
Because the microscopic state of polymer chains determines
the quality of the resultant polymer product,
it has been desired to develop a new simulation technique
that can predict not only the macroscopic flow behavior
but also the microscopic state of its constituent polymer chains.

For this purpose, 
an increasing amount of attention has recently been paid to a
multiscale simulation (MSS) method that can bridge the gap
between the macroscopic flow simulator
and a large number of molecular-based polymer dynamics 
simulators.
A pioneering study of multiscale simulations
is the Calculation of Non-Newtonian Flow: Finite Elements
and Stochastic Simulation Technique (CONNFFESSIT),
which was developed by Laso and {\"{O}}ttinger\cite{Laso1993}. 
In the CONNFFESSIT framework, 
macroscopic balance equations and a microscopic system
consisting of a dumbbell model are successfully combined.
As a review on multiscale approaches, the literature
by Keunings \cite{Keunings2004} will be useful.
Based on the CONNFFESSIT, various types of
multiscale simulation methods
have been proposed\cite{Hulsen1997, Halin1998, E2003, Yasuda2008, Murashima2010}. 

Thus far, we have succeeded in applying a multiscale simulation
method\cite{Murashima2010} to isothermal flow problems of well-entangled
monodispersed polymer melts in various flow geometries such as flows
around an infinitely long cylindrical obstacle\cite{Murashima2011, Murashima2012}, 
those in a 4:1:4 contraction-expansion channel\cite{Sato2019a},  
and those of the polymer melt spinning process\cite{Sato2017}.
In our multiscale simulation for well-entangled polymer melts,
a macroscopic flow simulation model for solving the balance equations is
combined with a number of microscopic systems described
by a slip-link model\cite{Doi2003}.
We conducted multiscale simulations of the systems mentioned above
under isothermal conditions, but industrial processes
are usually performed under nonisothermal conditions.
In addition, industrially used polymeric materials are not monodisperse but
polydisperse in the molecular weight.
These nonisothermal and polydisperse natures of polymeric systems
bring various novel flow behaviors and make the flow predictions difficult.
The multiscale simulation method used in the present work
can be applied to polydisperse systems by simply considering
poly-dispersed polymer chains in the slip-link simulator embedded
in each Lagrange particle.
Hence, the multiscale method would be a potential candidate to deal with such
a system.
In this work, however, to focus on and to demonstrate how the temperature
field affects the macroscopic flow and microscopic state of the polymer chains,
we restrict ourselves to a mono-dispersed, but nonisothermal polymeric
system.
We will tackle the system with polydispersity as future work.

It should be noted that multiscale simulations
under a nonisothermal condition
using the Kremer-Grest (KG) model\cite{Kremer1990} have been performed
by Yasuda and Yamamoto,
{\it i.e.}, the so-called synchronized molecular dynamics
(SMD)\cite{Yasuda2014}.
Using SMD, they successfully investigated polymer lubrication problems\cite{Yasuda2016,Yasuda2019}.
However, SMD is currently limited to simple flow problems 
because this method has, for example, a heavy computational cost.
A multiscale simulation technique that employs a more coarse-grained model
than the KG model is still desired to consider processing flow problems. 
Thus, the objective of the present paper is to develop a method to
incorporate the effect of the temperature distribution
into the multiscale simulation method
we have developed thus far.
The paper is organized as follows.
In Sec.~\ref{sec:model}, 
we explain the nonisothermal multiscale simulation method, 
especially how the temperature effect is taken into account
in both a macroscopic system and a microscopic system.
In Sec.~\ref{sec:Results and Discussion},
we show the results obtained in the application
of the extended method to a nonisothermal flow problem
between two coaxial cylinders.
Finally, in Sec.~\ref{sec:Conclusion},
we draw conclusions regarding the extended multiscale simulation method.
%
%
\begin{figure}[t]
\centering
\includegraphics[width=80mm]{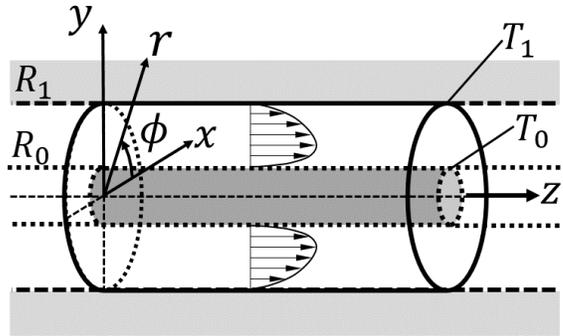}
 \caption{Schematic picture of a coaxial dual-cylinder system}
\label{fig:Fig01}
\end{figure}
%

\section{\label{sec:model}Model}

\subsection{Macroscopic System}

We consider a polymer melt flow between two infinitely long coaxial cylinders,
where the respective radii of the inner and outer cylinders are 
$R_0$ and $R_1$ (Fig.\ref{fig:Fig01}).
This type of flow geometry can be seen in manufacturing in polymer products
that have a cylindrical cavity along the center line of the cylindrical extrudates,
such as a tube, parison, and hollow fiber.
The flow problem considered here, however, is somewhat artificial.
Because the aim of the present paper is to demonstrate the extensibility
of the MSS method, originally developed for iso-thermal flow problems,
to nonisothermal flow of well-entangled polymer melts, 
it is important to make an assessment of the nonisothermal MSS method
by choosing a simpler flow geometry. Therefore,
we choose this infinitely long coaxial cylinder geometry.
To describe this system, we use a cylindrical coordinate system whose $z$-direction
is set to the center axis of the coaxial dual-cylinders, 
whose radial direction is perpendicular to the center axis
and whose azimuthal direction ($\phi$-direction) is defined 
as shown in Fig.\ref{fig:Fig01}. 
The flow between the coaxial cylinders is induced by applying a uniform pressure gradient.
The temperatures of the inner and outer cylinders are set to $T_0$
and $T_1$ ($T_0 \geq T_1$), respectively.
The viscosity of the polymer melt considered here is high enough
such that one can assume a laminar flow. 
For the sake of the translational symmetry of the system along the $z$-direction, 
all variables are functions of the radius $r$ and time $t$, 
and the velocity is expressed as $\boldsymbol{v}=v_z(r,t) \be_z$,
where $\be_z$ is the unit vector along the $z$-axis 
and the temperature is $T(r,t)$.
The equations for $v_z$ and $T$ are given by 
\begin{eqnarray}
&& \rho {\partial v_{z} (r,t) \over \partial t}
   = 
{1 \over r}{\partial \over \partial r}
 \Bigr ( r \sigma_{rz} ( r,t ) \Bigr )
 + F_{z}^{(\rm ext)}
 \label{TimeEvolution_V}
 \\
&&  \rho C_p \frac{\partial T(r,t)}{\partial t}
    =k \frac{1}{r}\frac{\partial}
   {\partial r}\left( r \frac{\partial T}{\partial r} \right )
   + \sigma_{rz} \frac{\partial v_z}{\partial r} 
\label{TimeEvolution_T}
\end{eqnarray}
where
$\rho$ is the density of the fluid, 
$\sigma_{rz}$ is the $rz$-component of the stress
coming from the entangled polymer dynamics, 
and $F_{z}^{(\rm ext)}$ is a constant force density given
by a constant pressure gradient along the $z$-direction.
In the energy equation \eqref{TimeEvolution_T}, 
$C_p$ and $k$ are the specific heat capacity and
thermal conductivity of the fluid, respectively.
The boundary conditions at the two walls for the velocity and
temperature
are respectively given as
\begin{eqnarray}
&&  v_{z} (R_0,t)=0, \quad v_{z} (R_1,t)=0,
\label{eqn:BC_for_v}
\\
&&    T(R_0,t)=T_0,  \quad T(R_1,t)=T_1.
\label{eqn:BC_for_T}
\end{eqnarray}
As the units of length, time and stress, we employ the radius of the outer cylinder $R_1$,
$\tau_\re(T_{\rm ref})$ and $\sigma_\re(T_{\rm ref})$, 
respectively, where $T_{\rm ref}$ denotes a reference temperature.
$\tau_\re(T_\rmref)$ and $\sigma_\re(T_\rmref)$
are also the units of time and stress for the microscopic model used in the present work,
which will be explained later. 
The velocity and the external force density
are scaled by the constructed units
$v_\ro\equiv R_1/ \tau_{\rm e}(T_{\rm ref})$
and
$F_\ro \equiv\sigma_{\rm e}(T_{\rm ref})/R_1$,
respectively.
Generally,
the density and the heat capacity of a polymer melt also depend on the temperature, 
but here, they are treated as constant because their variation with
the temperature is considered to be small.
In addition, the thermal conductivity $k$ may have a tensorial form $k_{\a\b}$
depending on a local polymer conformation that 
can be evaluated at the microscopic level in our multiscale simulation.
Because the amount of experimental data on the dependence of
$k_{\a\b}$ on the polymer conformation is insufficient,
the thermal conductivity is also simply assumed to be constant.
With the units mentioned above, 
eqs.\eqref{TimeEvolution_V} and \eqref{TimeEvolution_T}
become the following nondimensional expressions:
\begin{eqnarray}
{\Re^{\rm (a)}} \frac{ \partial {\tilde v}_{z}
 (\tilde{r}, \tilde{t})}{\partial \tilde{t}}
&=&\frac{1}{\tilde{r}} \frac{\partial}{\partial \tilde{r}}
\Bigl(\tilde{r} \tilde{\sigma}_{rz}
(\tilde{r}, \tilde{t}) \Bigr)+{\tilde{F}_z}^{\rm {(ext)}}
   \label{Eq_motion_nondim}
\\
 {\Pe^{\rm (a)}} \frac{ \partial T (\tilde{r}, \tilde{t})}{\partial \tilde{t}}
&=&
\frac{1}{\tilde{r}} \frac{\partial}{\partial \tilde{r}}
\left(\tilde{r} \frac{ \partial T }{\partial \tilde{r}} \right)+
{\Br^{\rm (a)}}\frac{ \partial \tilde{v}_z }{\partial \tilde{r}}\tilde{\sigma}_{rz}
\label{Eq_energy_nondim}
\end{eqnarray}
where the variables with the tilde symbol
on top of them are scaled variables.
$\Re^{\rm (a)}$, $\Pe^{\rm (a)}$ and $\Br^{\rm (a)}$
are the apparent Reynolds, P\'eclet and Brinkman numbers, respectively, and
are defined as
\begin{eqnarray}
\Re^{\rm (a)} ={\rho v_\ro R_1 \over \eta_\re(T_\rmref)}, 
\Pe^{\rm (a)} =\frac{v_\ro R_1}{D_T}, 
\Br^{\rm (a)} =\frac{\sigma_{\rm e}(T_{\rm ref}) v_\ro R_1}{k}~~~~
\end{eqnarray}
where $\eta_\re(T_\rmref)=\sigma_\re(T_\rmref) \tau_\re(T_\rmref)$ and $D_T=k/\rho C_p$.
The three parameters $\Re^{\rm (a)}$, $\Pe^{\rm (a)}$ and $\Br^{\rm (a)}$
that appear in the macroscopic equations \eqref{Eq_motion_nondim} and \eqref{Eq_energy_nondim}
are determined by the geometry of the system and the material considered here.

\subsection{Microscopic Model of Entangled Polymer Chains}
\label{subsec:micro}
As a microscopic model to describe the rheological properties
of a well-entangled polymer melt,
we employ the dual slip-link model proposed by Doi and
Takimoto\cite{Doi2003, Sato2019b}
but extend it such that it can describe the dynamics
of a well-entangled polymer melt under a time-dependent temperature $T(t)$.
Generally, 
it is not straightforward to incorporate
the temperature effect into a coarse-grained polymer
model, unlike an atomistic polymer model, 
because the temperature dependence of parameters
used in the coarse-grained model is not clear.
It is well known that the time-temperature superposition (TTS)
rule\cite{Ferry1980} holds for the rheological properties
of a homogeneous polymeric liquid,
and the temperature effect has been successfully incorporated into a
constitutive equation
by scaling its parameters based on 
this time-temperature superposition rule
\cite{Habla2012,Pol2014,Zatloukal2016,Zhuang2017,Gao2019}.
The slip-link model is specifically used to describe the rheological
properties of a homogeneous well-entangled polymer melt;
therefore, the TTS rule can be used inversely to determine
the temperature dependence of the parameters
used in the coarse-grained model, 
such as the Rouse and reptation relaxation times, 
and the temperature dependence of the stress. 
Namely, 
the way to incorporate the temperature effect
into the model is to use the TTS rule,
which means that all the physical quantities related to time
and stress must be scaled by shift factors $a_T$ and $b_T$,
respectively. 
The shift factors $a_T$ and $b_T$ are given by 
\begin{eqnarray}
\log_{10}{a_{T}}= - \frac{C_{1}(T-T_\rmref)}{C_{2}+(T-T_{\rm {ref}})},
 \quad
{b_T}=\frac{T}{T_{\rm ref}}  
\label{eqn:WLF}
\end{eqnarray}
where $T_\rmref$ is the reference temperature used
in the TTS rule, $C_1=8.86$ and $C_2=101.6$ K.  
Here, the expression for the shift factor 
$a_T$ is called the Williams-Landel-Ferry (WLF) equation. 
Based on this idea, 
we can extend the slip-link model proposed
for an isothermal polymer melt system to a nonisothermal system, 
where the time unit and the stress unit are 
scaled by the shift factors $a_T$ and $b_T$, respectively. 
By using the extended model, 
we can apply it to a multiscale simulation of a polymer melt flow
with a temperature distribution in a flow channel.
The polymer melt considered here is composed of linear polymer chains 
with a monodisperse molecular weight distribution
and with a total number of strands $Z_0$
(in other words, with entanglements $Z=Z_0 + 1$)
on a single chain
in equilibrium.

\begin{figure}[t]
\centering
\includegraphics[angle=-90,width=72.0mm]{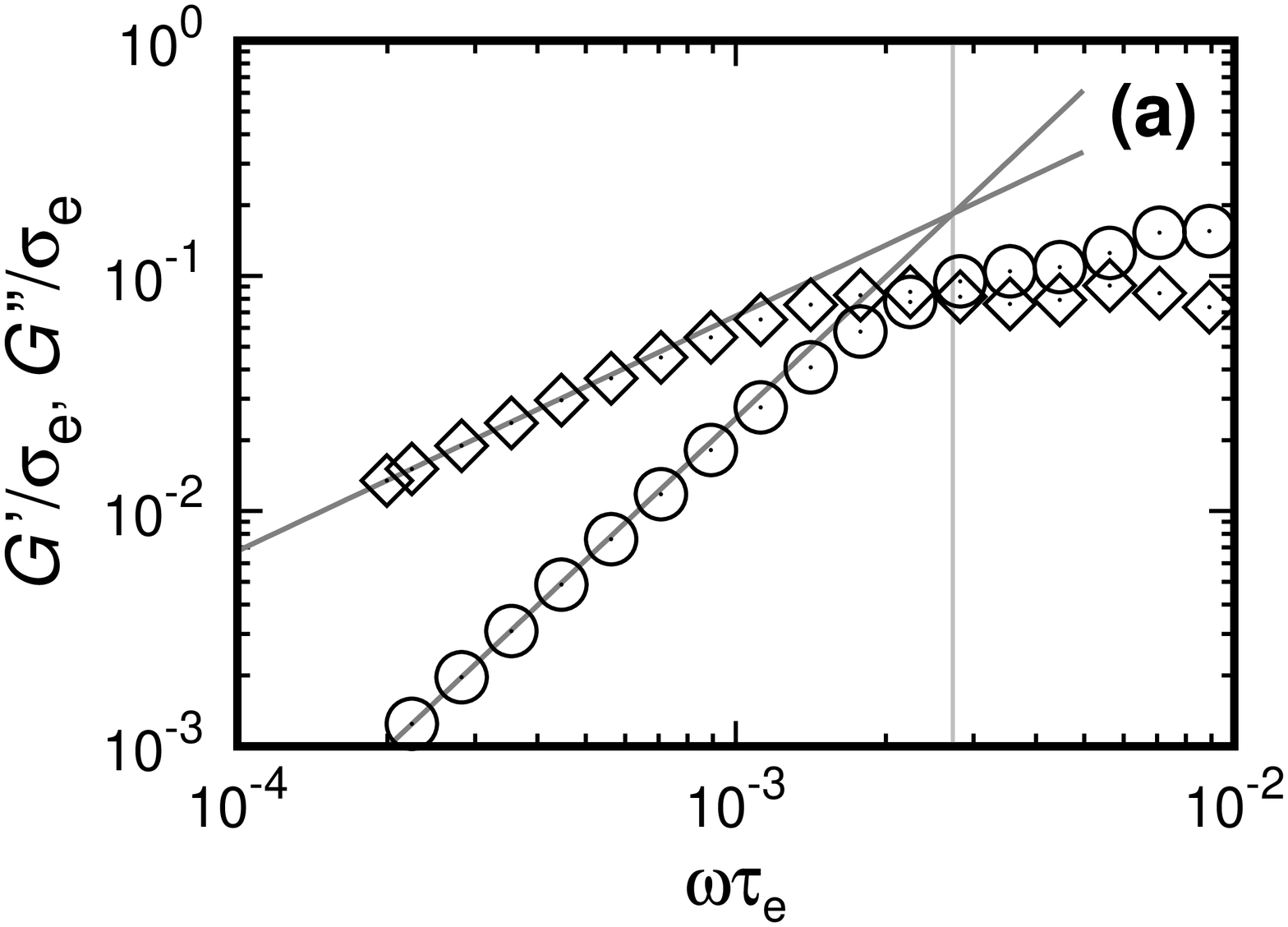}\\
\includegraphics[angle=-90,width=72.0mm]{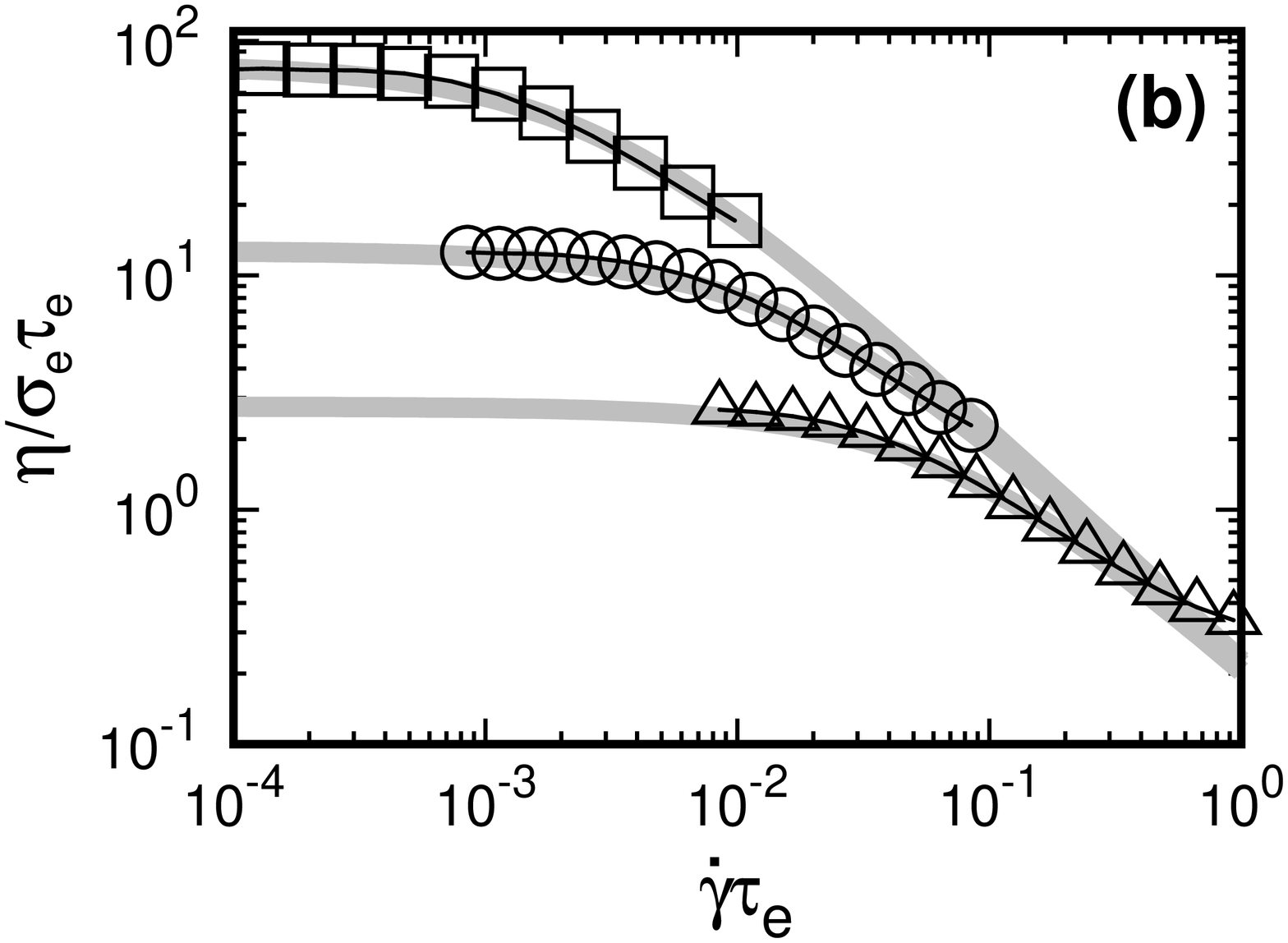}\\
\includegraphics[angle=-90,width=72.0mm]{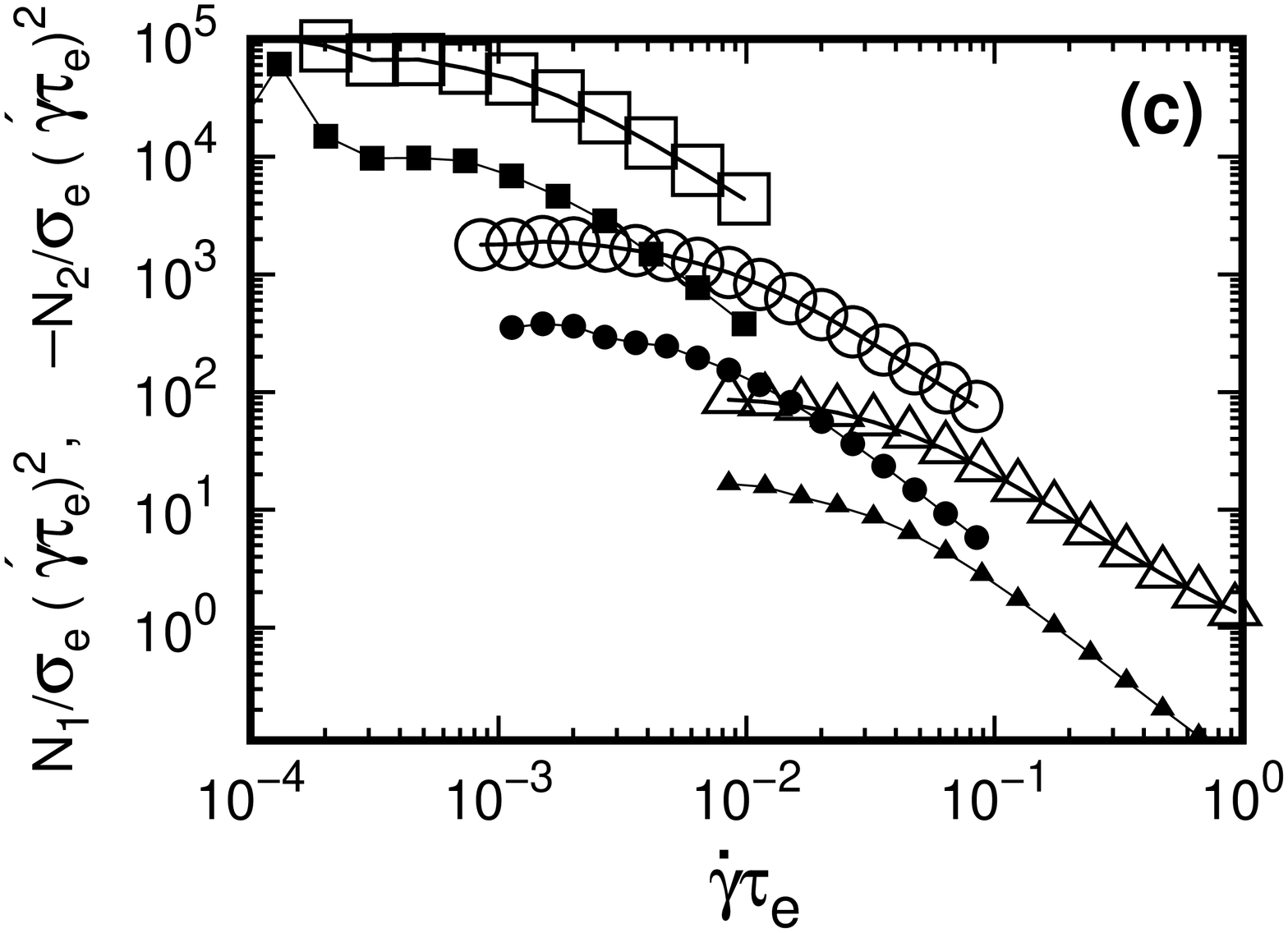} 
\caption{
 Rheological properties of the polymer melt ($Z_0=10$) used in the MSS.
 (a) linear viscoelasticity at $T=160^\circ$C ($\bigcirc$: $G'$,
 $\diamondsuit$: $G''$),
 (b) steady shear viscosity
     and
 (c)     1{\it st} normal stress difference
 (open symbol)
     and 2{\it nd} normal stress difference
 (filled symbol),
 at
 $T$=160$^\circ$C ($\square$),
     170$^\circ$C ($\bigcirc$),
 and 180$^\circ$C ($\bigtriangleup$) in (b) and (c).
In (b), the top gray line
is fitted to the data at 160$^\circ$C ($\square$) using eq.\eqref{eqn:Cross_WLF_model}, 
where the estimated fit parameters are 
$\eta_0\left(T_\rmref\right)=78.8$, $n=0.993$ and
$C_3=4.55$.
The remaining two gray lines are drawn by respectively
setting the temperature to 170$^\circ$C and 180$^\circ$C in
eq.\eqref{eqn:Cross_WLF_model},
with the estimated parameters unchanged. 
 }
  \label{fig:Fig02}
\end{figure}
%
%
In the slip-link model, 
a polymer chain is expressed by a primitive path and
slip-links on the primitive path; 
thus, the length of a primitive path in equilibrium is $L_0 = Z_0 a$,
with $a$ being the distance between two consecutive entanglements
along a chain in equilibrium. 
A slip-link represents an entanglement point between two different polymer chains.
The number of slip-links $Z$ on a primitive path changes with time $t$. 
The polymer chain expressed by a primitive path is composed of $Z$ slip-links, 
$(Z-1)$-strands and two tails. 
The primitive path length of a polymer chain is given by 
\begin{equation}
L = s_{\rm +} + s_{\rm -} + \sum_{j = 1}^{Z-1}|{\br_j}|
\label{eqn:primitive_path_length}
\end{equation}
where $s_+$ and $s_-$ are the lengths of the head and tail, respectively,
and ${\br}_j$ is the relative vector from
the position of the $j$-th slip-link to that of the $(j-1)$-th slip-link.
For details on the procedure for updating the state of a primitive path
and the slip-links of a polymer in a fluid particle, see Appendix I. 
%
%
%
\begin{figure}[t]
\centering
\includegraphics[angle=0,width=75mm]{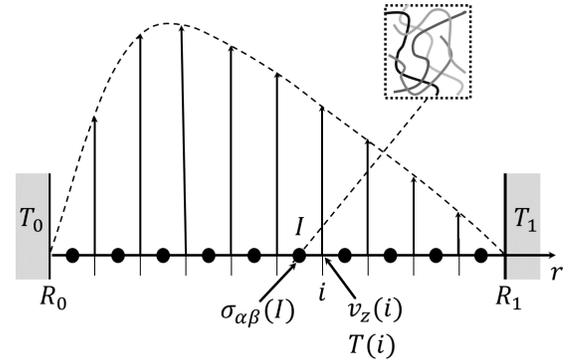}
\caption{
Schematic image of the discretization of the system along the radial direction.
}
\label{fig:Fig03}
\end{figure}

\begin{figure}[t]
 \centering
\includegraphics[angle=-90,width=80mm]{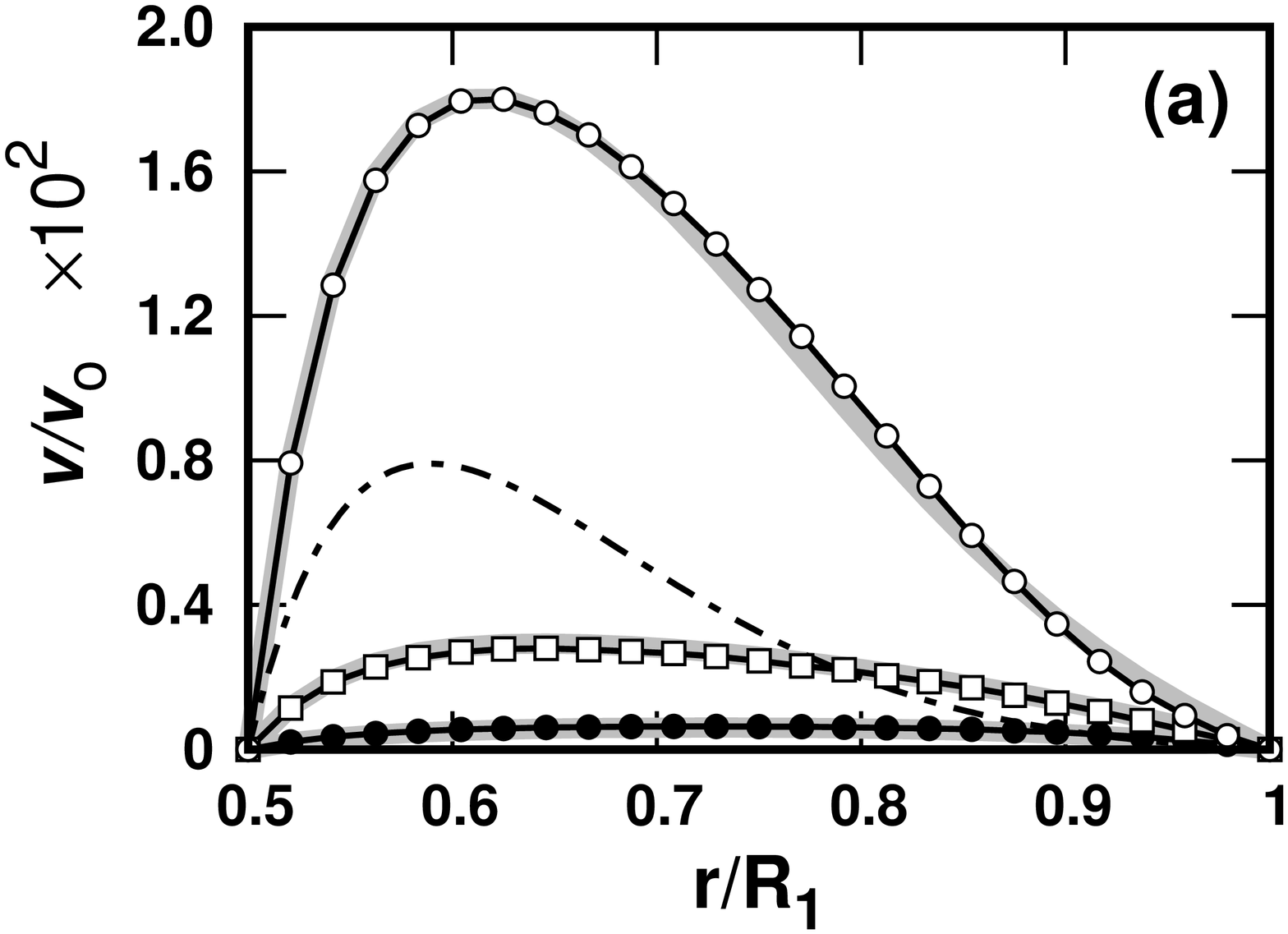}
\includegraphics[angle=-90,width=80mm]{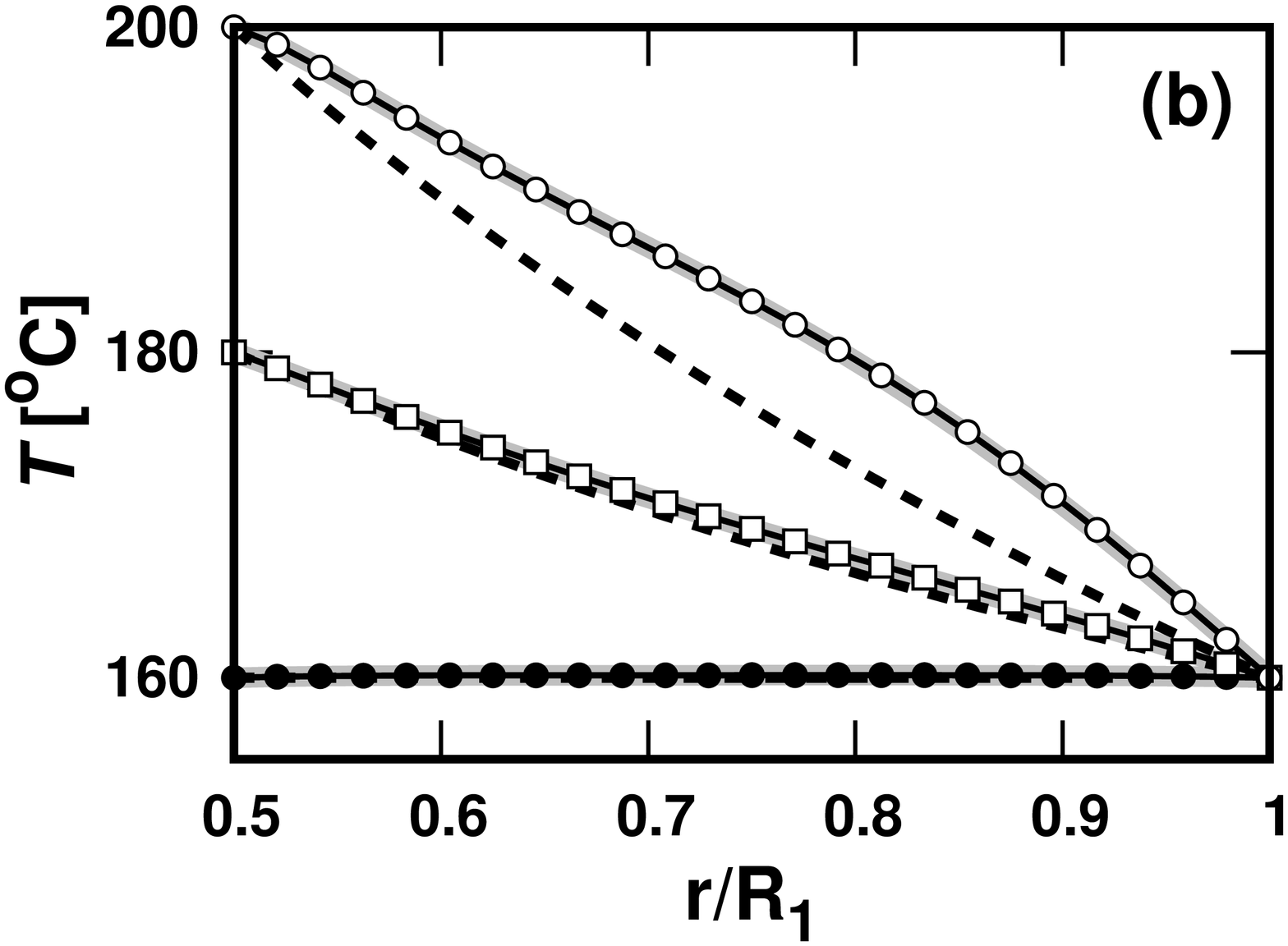}
\caption{
(a) Velocity and (b) temperature profiles
 along the radial direction at steady states for
$T_0$ equal to (i) 160$^\circ$C (filled circle),
              (ii) 180$^\circ$C (open square),
         and (iii) 200$^\circ$C (open circle).
 In (a), the
 dash-dotted
 line represents condition (iii) but with a Newtonian
 fluid with zero shear viscosity $\eta_0(T)$
 $=\eta_0(T_{\rm ref}) a_T b_T$
 that is equivalent to that
 of the slip-link model ($Z_0=10$).
 In (b), the dashed line just below each solid line
 stand for the analytical solution to eq.\eqref{Eq_energy_nondim}
 with $\Br^{\rm (a)}=0$ and $\Pe^{\rm (a)}=0$.
In (a) and (b), the three thick gray lines underlying
the simulations data (filled and open symbols)
are obtained from eqs.\eqref{Eq_motion_nondim} and \eqref{Eq_energy_nondim},
with the stress given by the Cross-WLF model of
eq.\eqref{eqn:Cross_WLF_model}, 
using the estimated fit parameters of Fig.\ref{fig:Fig02}.
}
\label{fig:Fig04}
\end{figure}
\begin{figure}[t]
 \centering
 \includegraphics[angle=-90,width=80mm]{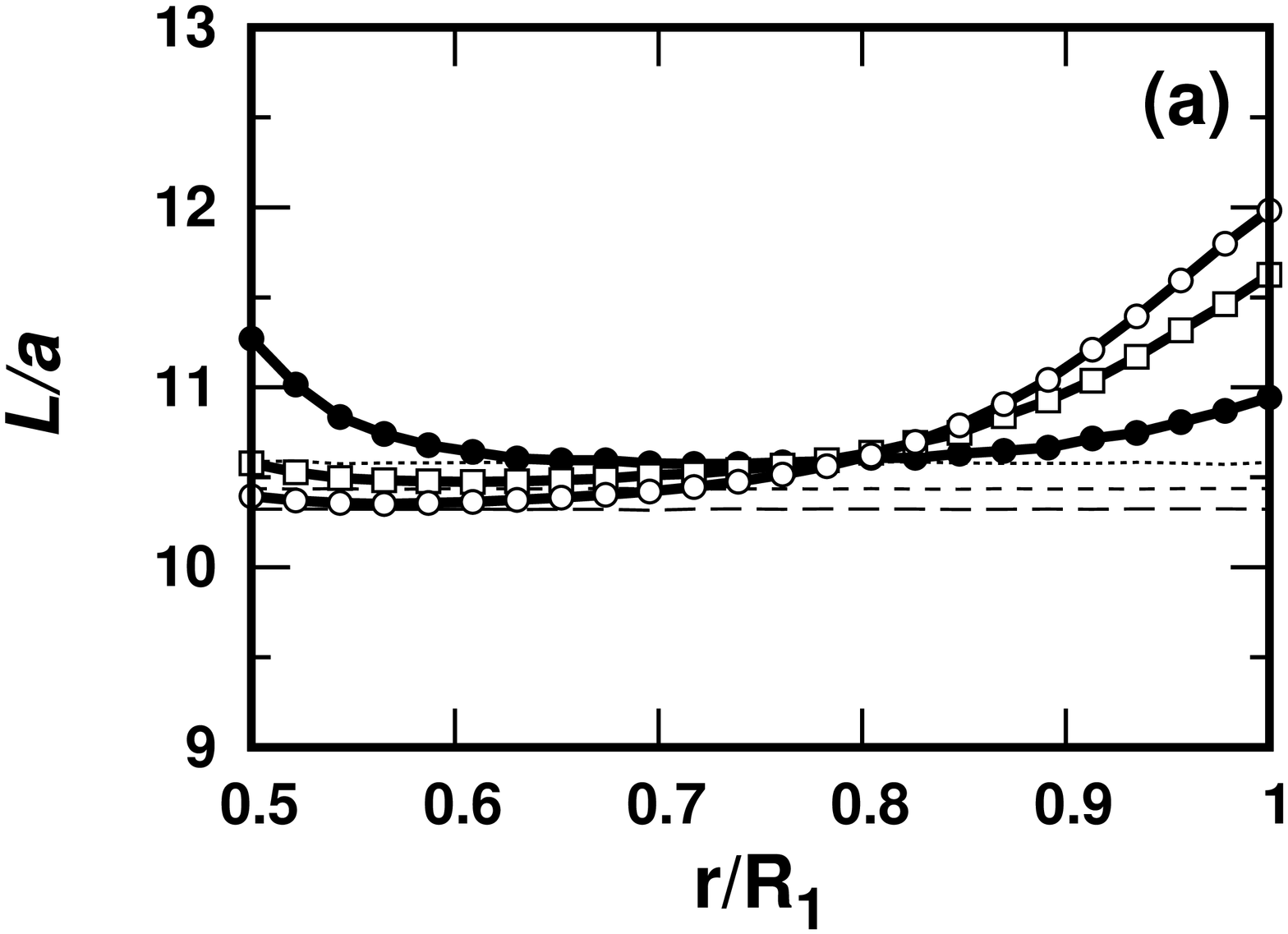}
 \includegraphics[angle=-90,width=80mm]{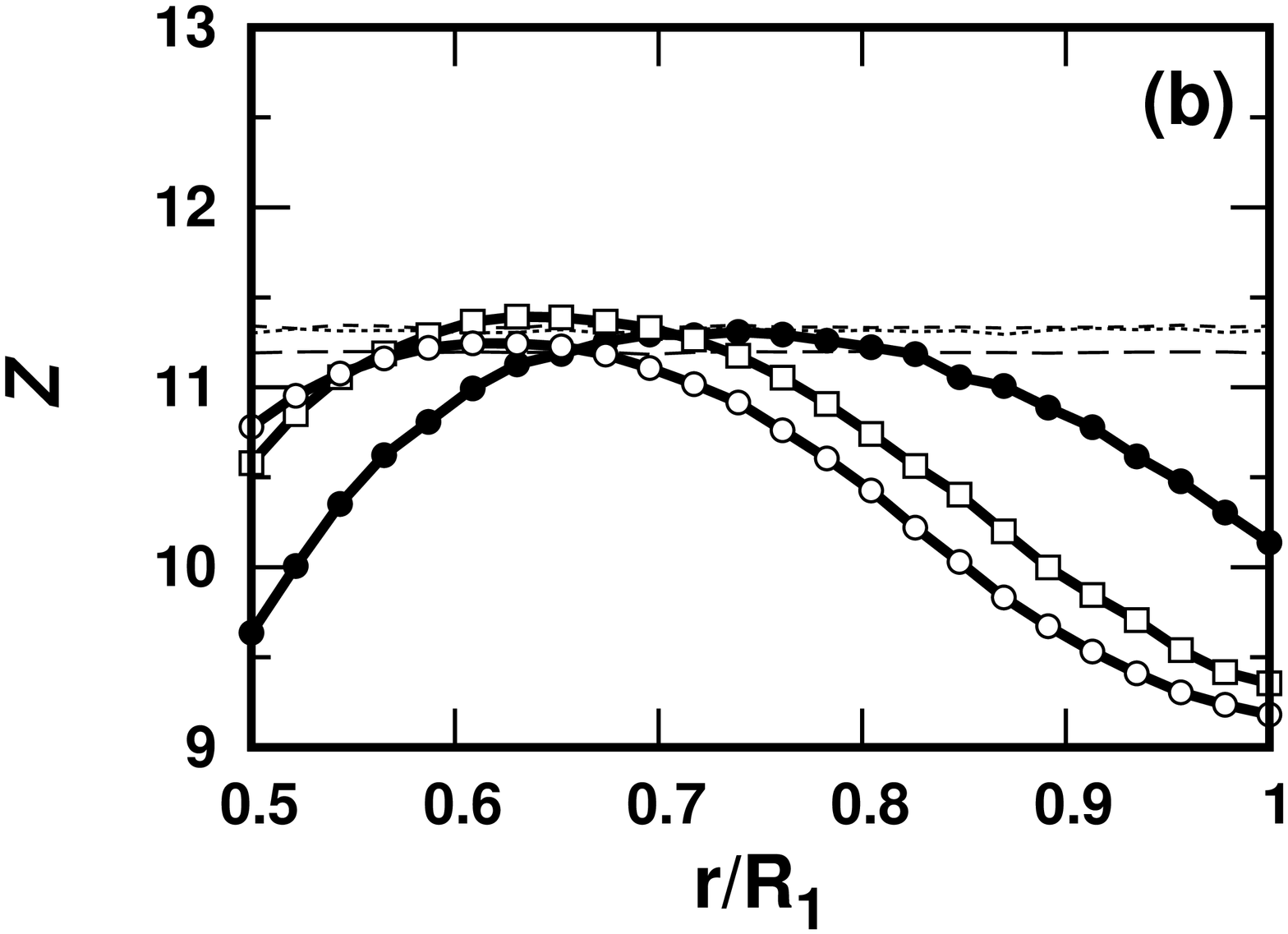}
\caption{
  (a) Average length of primitive paths and
  (b) average number of entanglements 
  on a primitive path 
  as functions of the radius at the steady states
  for
$T_0$ equal to (i) 160$^\circ$C (filled circle),
      (ii) 180$^\circ$C (open square),
 and (iii) 200$^\circ$C (open circle).
  The dotted, dashed, and long-dashed lines stand for the data
  in equilibrium at $T_0=$160, 180, and 200$^\circ$C, respectively.
  }
\label{fig:Fig05}
\end{figure}

  \subsection{
  Bulk Rheological Properties of a Polymer Melt}
  \label{subsec:Bulk Rheological_Properties_of_a_Polymer_Melt}
As the polymer melt in our simulation, 
we consider a system of monodispersed entangled linear polystyrene (PS). 
Each of the constituent polymers has
ten entangled strands $Z_0=10$ ({\it i.e.}, eleven entanglement points, $Z=11$) 
in equilibrium,
which correspond to a polystyrene with a molecular weight of $1.4 \times
10^5$.
We first investigate
linear viscoelastic properties and the bulk steady shear viscosity 
of the polymer melt,
which are used in our multiscale simulations
in Sec.~\ref{sec:Results and Discussion}. 
In Fig.\ref{fig:Fig02}(a), we show 
the storage modulus $G'$ and loss modulus $G''$ evaluated from
the stress relaxation modulus $G(t)$
after applying a step shear strain with $\gamma_0=0.5$
to the entangled polymer melt at $T=160^\circ$C.
The reptation time $\tau_\rd$
is found to be $366 \tau_\re$ at $160^{\rm o}$C 
from the intersection of the two lines (gray lines in
Fig.\ref{fig:Fig02}(a)) fitted to $G'$ and $G''$ in the terminal relaxation region.
Fig.\ref{fig:Fig02}(b) shows
the steady shear viscosities of the previously mentioned polymer melt 
at three temperatures: 160$^\circ$C, 180$^\circ$C and 200$^\circ$C. 
As seen from Fig.\ref{fig:Fig02}(b), 
the systems exhibit the shear thinning behavior.
To analyze the temperature dependent shear thinning behavior,
we used the following Cross-WLF model, defined as
\begin{equation}
 \eta(\dot \gamma, T) = {\eta_0(T_\rmref) a_T b_T \over 1 + C_3
  (\eta_0(T_\rmref) a_T |\dot \gamma| )^n}
  \label{eqn:Cross_WLF_model}
\end{equation}
where $\eta_0(T_\rmref)$ is the zero shear viscosity of the polymer melt
at $T_\rmref$, $C_3$ and the exponent $n$ are fitting parameters.
By fitting the data obtained by the slip-link model at
$T_\rmref=160^\circ$C 
with the Cross-WLF model (eq.\eqref{eqn:Cross_WLF_model}),
the parameters are found to be  $\eta_0\left(T_\rmref\right)=78.8$,
$n=0.993$ and  $C_3=4.55$.
In Fig.\ref{fig:Fig02}(b), the fitted line for
$T_\rmref=160^\circ$C,
together with the corresponding lines for $T=180^\circ$C and $T=200^\circ$C,
evaluated using the same parameter values, are drawn using thick gray lines.
One can clearly see that the temperature dependent shear thinning
behaviors are captured by the Cross-WLF model.

\subsection{
  Multiscale Simulation Method
  }
In the present multiscale simulation,
the strain rate tensor field evaluated at the macroscopic level
and the stress tensor evaluated at the microscopic level
are connected.
The region between the two coaxial cylinders
is divided along the radial direction into $M$-fluid elements
with an interval $\D \tilde r =(R_1-R_0)/M R_1$
and $(M+1)$-mesh points at the macroscopic level, as shown in
Fig.\ref{fig:Fig03}.
Each fluid particle contains $N_\rp$-polymer chains .
In this work, we used $M=24$ and $N_\rp=10^4$.
For the sake of symmetry of the system, 
all physical variables depend only on $r$
and $t$. 
The velocity and temperature are evaluated at every lattice point
by eqs.\eqref{Eq_motion_nondim} and \eqref{Eq_energy_nondim}, 
while the stress tensor is evaluated
at the every mid-point between two adjacent lattice points 
by using the microscopic model described in Sec.~\ref{subsec:micro}. 
%
%
%
%
\section{
\label{sec:Results and Discussion}Results and Discussion}
By using the following physical quantities related to a polystyrene melt
with $Z_0=10$ at $T_\rmref=160^\circ$C as
$\tau_{\rm e}(T_{\rm ref})=2.2$ msec,
$\sigma_{\rm e}(T_{\rmref})=0.5$ MPa,
$\rho=1040$ kg/${\rm m}^3$,
$C_p=1320$ J/(kg$\cdot$K), 
$k=0.136$ J/(m$\cdot$s$\cdot$K),
and 
$D_T=9.9 \times 10^{-8}$ m$^2$/s
and the quantities related
to the process parameters
$R_1=5.0$ mm, $R_0=2.5$ mm, 
$v_0=2.3$ m/s and $F_\ro^{\rm (ext)}=1.0 \times 10^2$ MPa/m, 
the three nondimensional parameters that appear
in eqs.\eqref{Eq_motion_nondim} and \eqref{Eq_energy_nondim}
are evaluated to be 
$\Re^{\rm (a)}=1.1 \times 10^{-2}$, 
$\Pe^{\rm (a)}=1.1 \times 10^5$, and
$\Br^{\rm (a)}=4.2 \times 10^4$.
The temperature at the outer cylinder surface is fixed at $T_1=160^\circ$C throughout the present paper. 
We investigate the following three cases of the temperature at the inner cylinder surface
$T_0$: (i) 160$^\circ$C,
      (ii) 180$^\circ$C,
 and (iii) 200$^\circ$C.
\subsection{
\label{sec:steady state}
  Steady state under an external force density along the $z$-direction
  }

In this subsection we focus on steady states under
the external force density $F_{z}^{\rm (ext)}=0.65 F_\ro$, 
where only the Brinkman number $\Br^{\rm (a)}$ is an important physical constant, 
as seen from eqs.\eqref{Eq_motion_nondim} and \eqref{Eq_energy_nondim}. 
The symbols in Fig.\ref{fig:Fig04}(a) and (b) show the velocity and temperature profiles
for (i)-(iii) at the steady state, respectively. 
In (a) and (b), the three thick gray lines underlying the simulations
data (filled and open symbols) are obtained from
eqs.\eqref{Eq_motion_nondim} and \eqref{Eq_energy_nondim},
with the stress given by the Cross-WLF model of
eq.\eqref{eqn:Cross_WLF_model}, 
using the estimated fit parameters of Fig.\ref{fig:Fig02}.
We can see from Fig.\ref{fig:Fig04}(a) that
the velocity becomes larger as $T_0$ increases.
This is because 
the Rouse and reptation relaxation times become exponentially shorter 
with increasing temperature; thereby, the viscosities become smaller.
The dash-dotted line in Fig.\ref{fig:Fig04}(a) represents the velocity profile of
a Newtonian fluid under the same boundary condition for $T$ as (iii)
but with a shear viscosity $\tilde \eta(T) = a_T b_T \tilde \eta_0(T_\rmref)$, 
the zero shear viscosity of which is the same as that of the polymer
melt ($Z_0=10$).
We can clearly see that the velocity for (iii) of the polymer melt
is larger than that of the Newtonian fluid.
Hence, the difference in velocity
is attributed to the shear thinning effect of the polymer melt, 
which can be seen in Fig.\ref{fig:Fig02}(b).
The dashed line in Fig.\ref{fig:Fig04}(b) is the analytic solution 
to the steady state of eq.\eqref{Eq_energy_nondim}
for $\Br^{\rm (a)}=0$, {\it i.e.}, for no heat generation, 
which is described as 
\begin{equation}
  T(r) = T_0+(T_1-T_0){\ln(r) - \ln(R_0) \over \ln(R_1)-\ln(R_0)}.
\label{eqn:analytic_solution_of_T}
\end{equation}
\noindent
We can clearly see that the temperature profiles for (i) and (ii) 
are almost the same as those of the analytical solution
to eq.\eqref{eqn:analytic_solution_of_T}.
However, the temperature profile in (iii) is larger than that
of the analytical solution without the heat generation
coming from the local shear stress and the shear rate. 
The deviation of the temperature profile from that given
by eq.\eqref{eqn:analytic_solution_of_T} is found to
originate from the larger velocity gradient, 
{\it i.e.,} the higher heat generation in (iii). 
Fig.\ref{fig:Fig05}(a) and (b) show
the distributions of the average length of a primitive path $L/a$
and the average number of entanglements on a chain $Z$, respectively.
The dotted, short-dashed and long-dashed lines in Fig.\ref{fig:Fig05}(a) and (b)
represent the values in the steady state
for the three cases (i)-(iii), respectively.
From Fig.\ref{fig:Fig05}(a), 
we can see that the polymer chains are stretched
at the regions near the outer cylinder wall,
and the stretching increases 
with the magnitude of the local velocity gradient. 
In the region close to the inner cylinder,  
on the other hand, the tendency is totally opposite.
Actually, the magnitude of the shear rate becomes larger with the local
temperature.
Nevertheless, the average length of the primitive path for (ii) and
(iii) is almost equal to those in equilibrium.
Namely, in this region,
the key factor in determining the polymer orientation
and stretch is not the local velocity gradient
but the local-temperature-dependent 
relaxation times $\tau_{\rm d}(T)$ and $\tau_{\rm R}(T)$.
Because the relaxation time becomes shorter with increasing temperature,
the chain can relax even at a high shear rate in the case of (ii) and (iii).
Similar to $L$ in Fig.\ref{fig:Fig05}(a), 
the behavior of $Z$ shown in Fig.\ref{fig:Fig05}(b) 
can be understood through
the convective constraint release (CCR) effect\cite{Marrucci1996}
with the local velocity gradient and 
temperature-dependent relaxation times.

\begin{figure*}[t]
\centering
\includegraphics[angle=-90,width=145truemm]{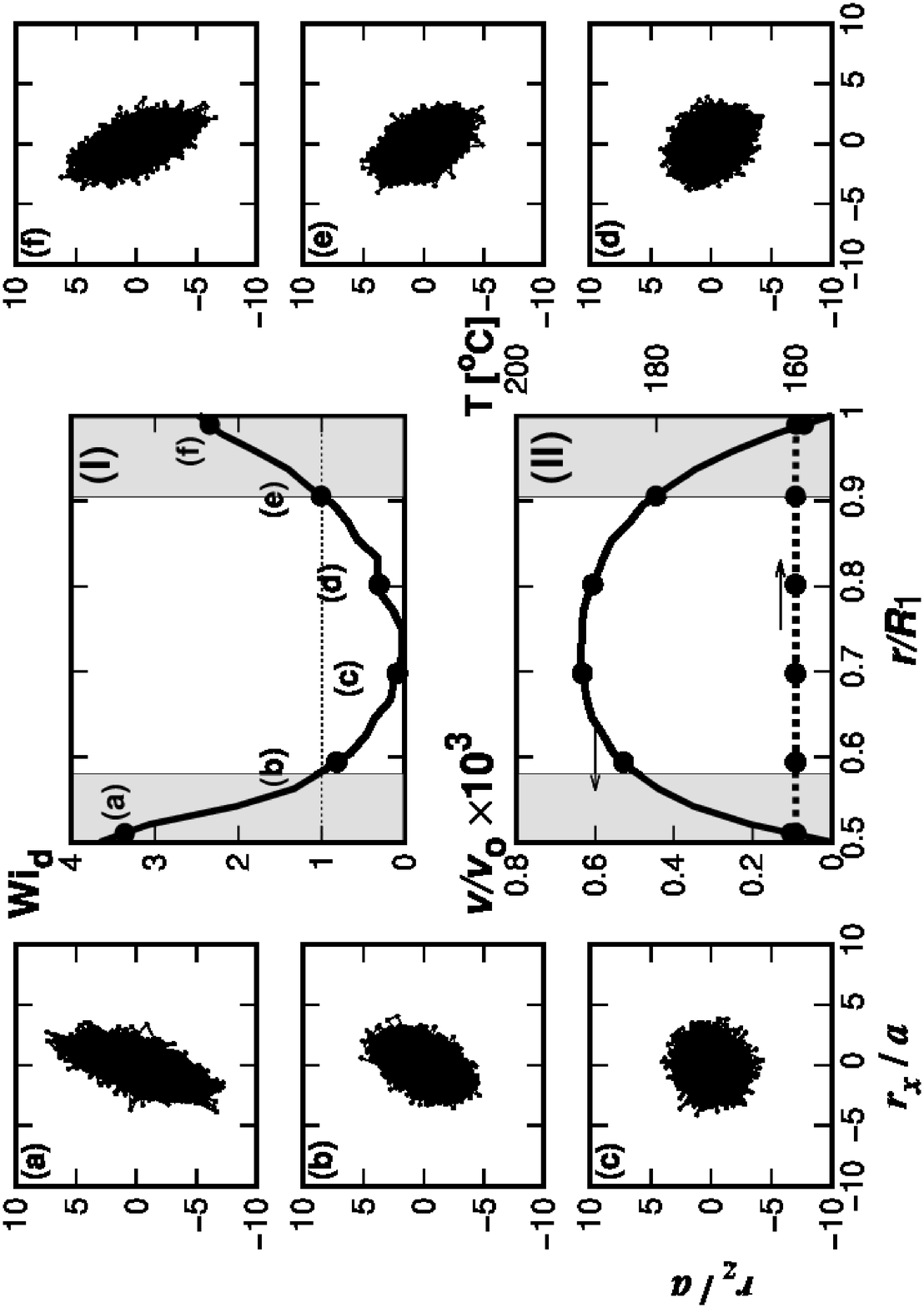}
\caption{
 (I) Weissenberg number profile $\Wi_\rd(r,T)$
 defined by eq.\eqref{eqn:WeissenbergNumber}
 and (II) velocity and temperature profiles as functions of the radius
 at the steady state for
 (i) $T_0=160^\circ$C.
 The figures (a)-(f) are the superimposed images of polymer conformations
 in the fluid elements located at $\tilde r=$
 (a) 0.52, 
 (b) 0.60, 
 (c) 0.71, 
 (d) 0.81, 
 (e) 0.92, 
 and 
 (f) 0.98, 
 where the primitive paths are drawn by fixing the center of mass 
 to the origin. The shaded regions in (I) and (II) represent
 the ones in which $\Wi_\rd > 1$.
}
\label{fig:Fig06:160}
\end{figure*}

\begin{figure*}[t]
\centering
\includegraphics[angle=-90,width=145truemm]{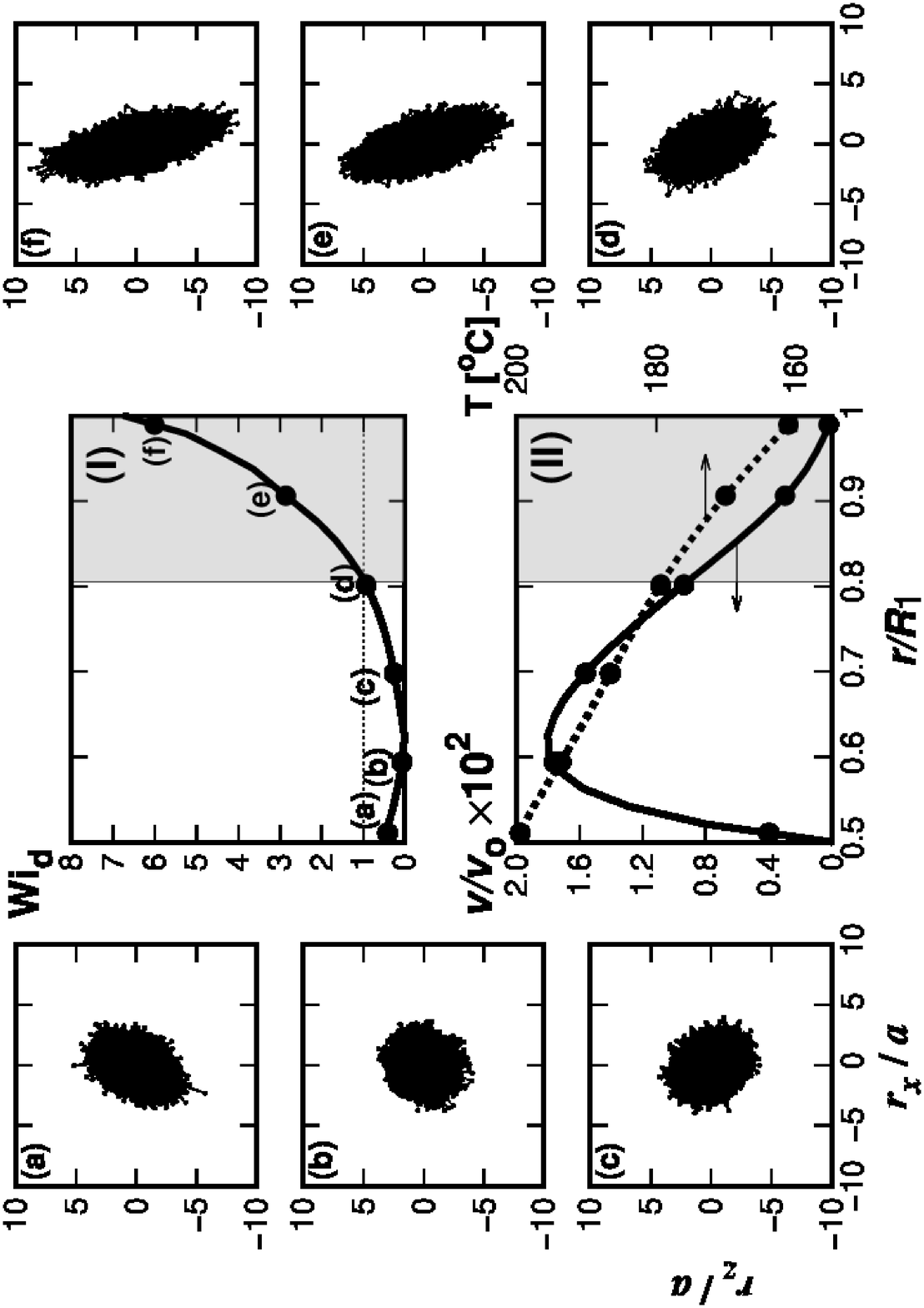}
\caption{
 (I) Weissenberg number profile
 $\Wi_\rd(r,T)$ defined by eq.\eqref{eqn:WeissenbergNumber}
 and (II) velocity and temperature profiles as functions of the radius
 at the steady state for (iii) $T_0=200^\circ $C.
 The figures (a)-(f) are the superimposed images of polymer conformations
 in the fluid elements located at $\tilde r=$
 (a) 0.52, 
 (b) 0.60, 
 (c) 0.71, 
 (d) 0.81, 
 (e) 0.92, 
 and 
 (f) 0.98, 
 where the primitive paths are drawn by fixing the center of mass 
 to the origin. The shaded regions in (I) and (II) represent
 the ones in which $\Wi_\rd > 1$.
}
\label{fig:Fig07:200}
\end{figure*}

Here, we focus on the relation between conformations of the polymer chains and
the reptation-time-based Weissenberg number $\Wi_\rd$, defined as
\begin{equation}
  \Wi_\rd(r,T) = \tau_\rd(T(r))
   \Bigr |{ \partial v_z \over \partial r }
   \Bigr |.
\label{eqn:WeissenbergNumber}
\end{equation}
$\Wi_\rd$ is expected to help guide us 
in considering how the extended polymer chain is oriented.
If $\Wi_\rd$ is larger than unity,
the polymer chains are expected to be oriented
along the flow direction.
In Fig.\ref{fig:Fig06:160} and Fig.\ref{fig:Fig07:200},
the Weissenberg number $\Wi_\rd$ 
is plotted as a function of $r$ 
at steady-state flows for the two typical cases of (i) and (iii), respectively.
Fig.\ref{fig:Fig06:160} and Fig.\ref{fig:Fig07:200} (a)-(f) also show 
conformations of the polymer chains expressed by primitive paths
in fluid elements located at the six typical points
in the tube for the cases of (i) and (iii) at the steady state.
In addition, 
the velocity and temperature distribution are shown
to visualize the relation between them and polymer conformations.
The shaded regions in Fig.\ref{fig:Fig06:160} and Fig.\ref{fig:Fig07:200} (I) and (II)
are the regions where $\Wi_\rd>1$, 
in which chains are oriented along the $z$-direction by shear flows. 
The states of polymer chains at the six typical points are drawn
in Fig.\ref{fig:Fig07:200}(a)-(f) as in Fig.\ref{fig:Fig06:160}. 
As shown in Fig.\ref{fig:Fig07:200} (I) and (II),
because $\tau_\rd$ becomes shorter as the temperature increases near the inner cylinder, 
the Weissenberg number becomes less than unity.
Namely, in the high temperature region close to the inner cylinder,
the relaxation rate of the chain orientation becomes faster than the shear rate in this region.
Although the polymer chains in this region experience a high shear rate, 
the polymer chains are not oriented that much along the flow direction
due to the Weissenberg number being less than unity in the above region. 
On the other hand, 
in the low temperature region close to the outer cylinder,
the relaxation rate is relatively slow, and the Weissenberg number is
larger than unity;
therefore, the chains are oriented along the flow direction. 
From the results shown in Fig.\ref{fig:Fig06:160} and Fig.\ref{fig:Fig07:200},
we can clearly see that the polymer chains are not necessarily oriented even at a high shear rate
depending on the temperature condition.
Namely, 
it can be said that the reptation-time-based Weissenberg number
$\Wi_\rd(r,T)$ is an effective measure
for considering a local orientation of polymer chains, especially in the nonisothermal case.
\begin{figure}[t]
\centering
\includegraphics[angle=-90,width=85truemm]{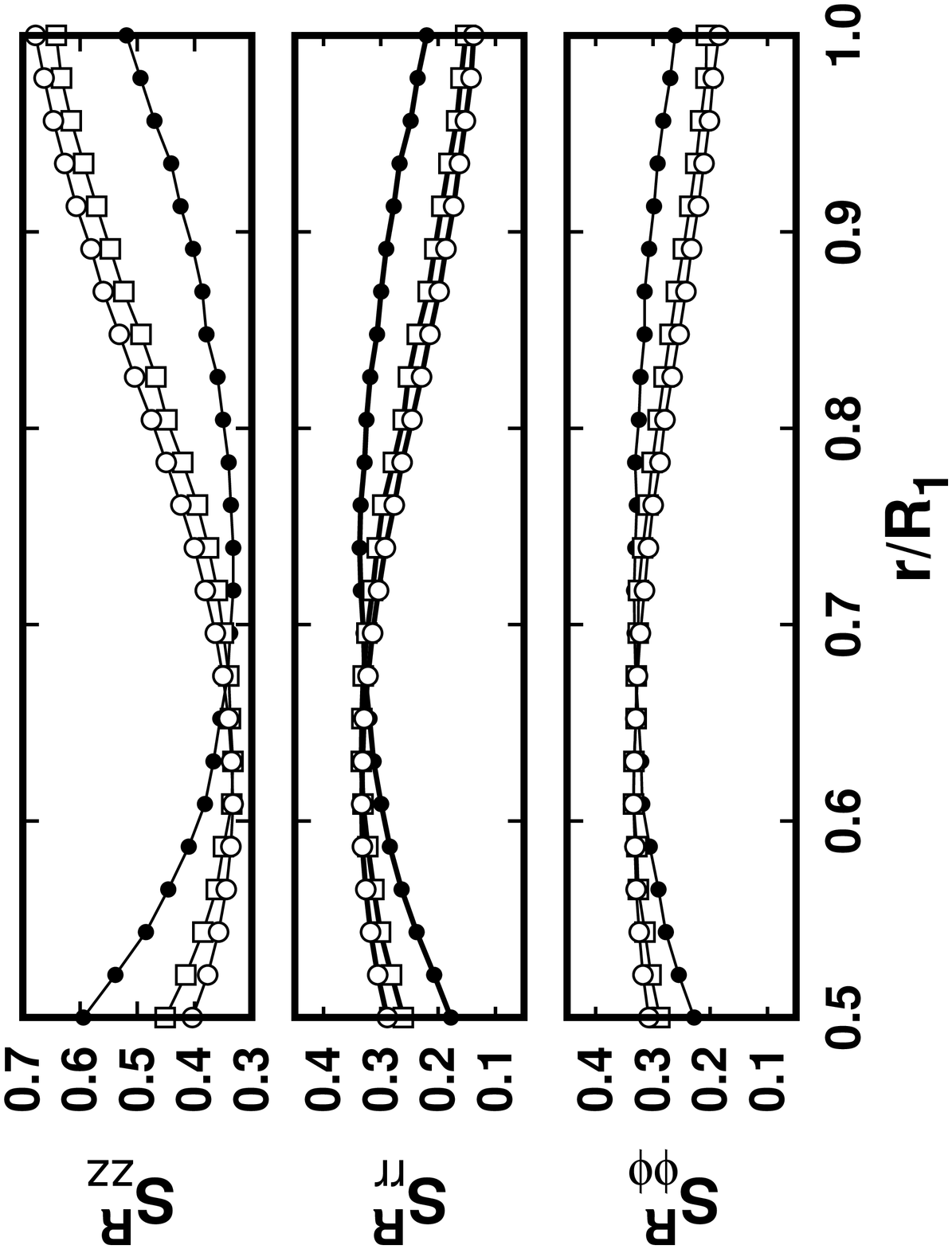}
\includegraphics[angle=-90,width=85truemm]{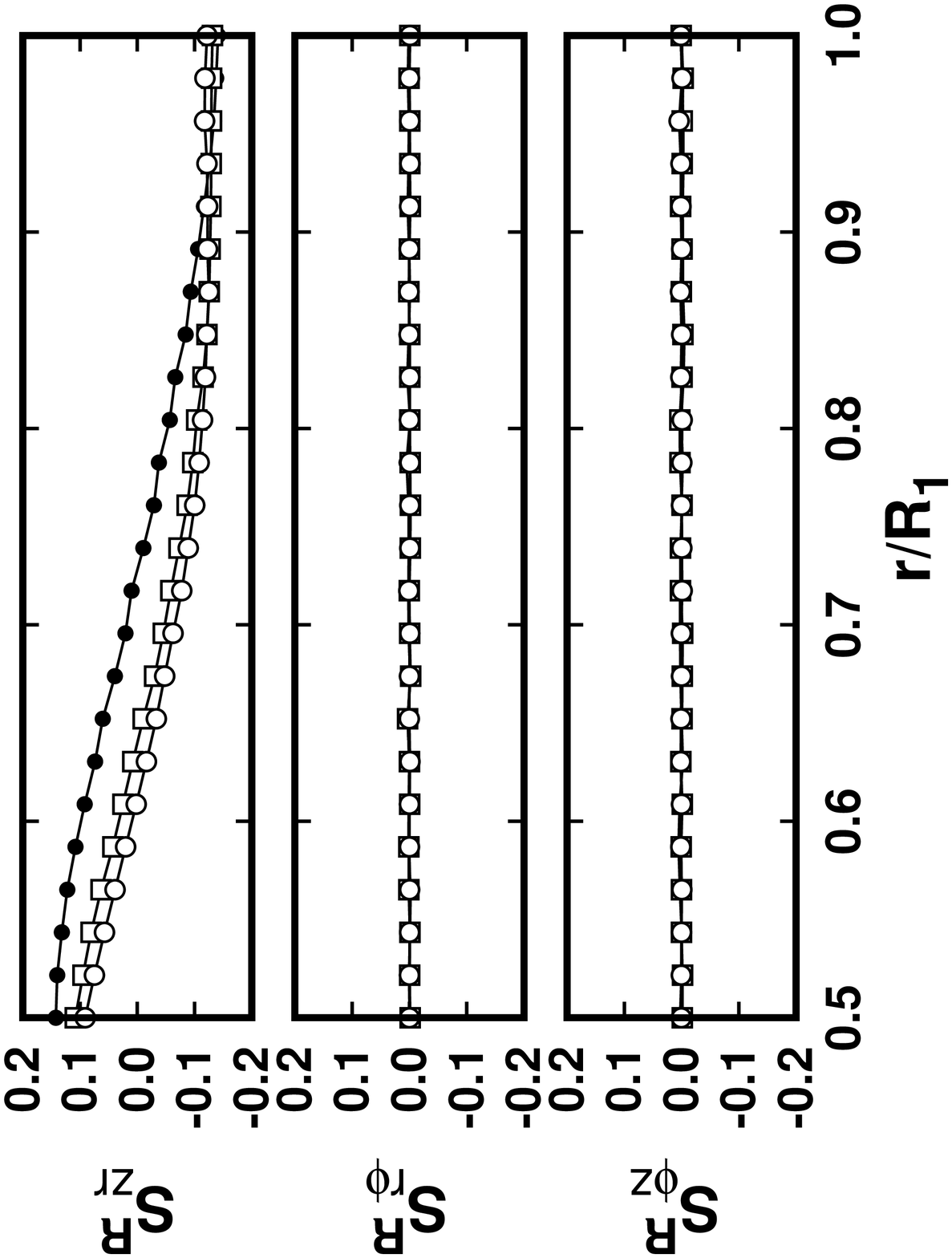}
\caption{
 End-to-end tensor $S_{\a\b}^{\rR}$ 
 ($\a,\b \in \{z, r, \phi \}$) as a function of $r/R_1$
  for
  $T_0$ equal to (i) 160$^\circ$C (filled circle),
                (ii) 180$^\circ$C (open square),
           and (iii) 200$^\circ$C (open circle).
 }
\label{fig:Fig08}
\end{figure}
\begin{figure}[h]
\includegraphics[angle=-90,width=85truemm]{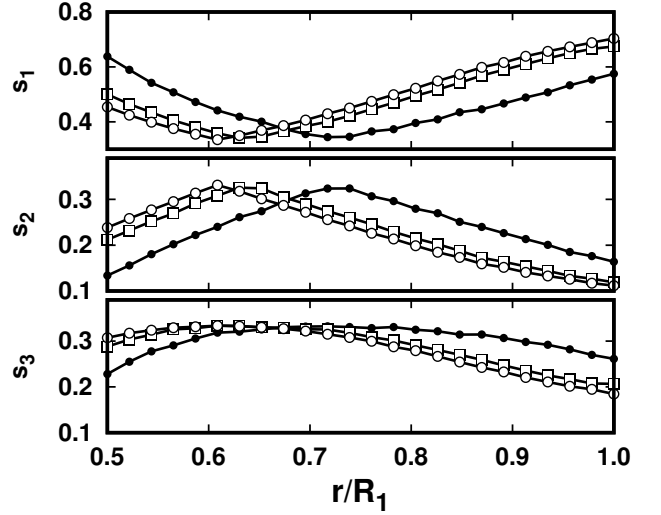}
\caption{
 Eigenvalues of $\bm S^{\rR}$
  for
  $T_0$ equal to (i) 160$^\circ$C (filled circle),
                (ii) 180$^\circ$C (open square),
           and (iii) 200$^\circ$C (open circle).
  $s_1$, $s_2$ and $s_3$
  are the eigenvalues whose 
  eigenvectors are almost along the $z$-, 
  $r$- and $\phi$-directions, respectively.
 }
\label{fig:Fig09}
\end{figure}
%
%
To quantitatively evaluate the microscopic states of polymer chains, 
we introduce the end-to-end tensor $\bm S^{\rR}$, defined as 
\begin{equation}
  {\bm S}^{\rR} = \biggr \langle {1 \over N_{\rm p} }
 \sum_{n=1}^{N_{\rm p}}
{
  {\bm R}^{(n)}  {\bm R}^{(n)}
 \over
 |{\bm R}^{(n)}|^2
}
\biggr \rangle
\label{eqn:conformation_tensor}
\end{equation}
where ${\bm R}^{(n)}$ is the end-to-end vector of a polymer chain $n$
and $N_{\rm p}$ is the total number of polymer chains.
$\langle (\cdots) \rangle$ denotes the ensemble average of $(\cdots)$, 
which is actually performed by taking a time average after the system
reaches the steady state.
Each component of the end-to-end tensor $\bm S^{\rR}$
in Fig.\ref{fig:Fig08} and eigenvalues 
of the tensor $\bm S^{\rR}$ in Fig.\ref{fig:Fig09}
are plotted as functions of the radial position $r$
for the three cases (i)-(iii).
Although $S_{zz}^{\rR}$, $S_{rr}^{\rR}$ or $S_{\phi\phi}^{\rR}$
can be described by the other two because of the relation
$S_{zz}^{\rR}+S_{rr}^{\rR}+S_{\phi\phi}^{\rR}$
$=1$,
all three are plotted for the purpose of comprehensibility.
The tensor $\bm S^{\rR}$ is ${\bm 1}/3$ in equilibrium, with $\bm 1$
being the unit tensor.
If $S_{\a\a}^{\rR}$  is larger than $1/3$, 
then the polymer chains are oriented along the $\a$-direction,
but if it is less than $1/3$, then the chains are less-oriented
along that direction.
Because the eigenvectors to the eigenvalues $s_1$, $s_2$ and $s_3$
are almost along the $z$-, $r$- and $\phi$-directions, respectively,
in similar way,
$s_\a> 1/3$ means that chains are oriented along the $\a$-direction.
As seen from Fig.\ref{fig:Fig08} and Fig.\ref{fig:Fig09}, $S_{zz}^{\rR}$
and $s_1$ are larger than 1/3, and therefore, the chains are highly
oriented along the $z$-direction (flow direction), 
especially at places close to the walls. 
The tendency $s_1$ near a wall of the outer cylinder
for the three cases can be understood by the velocity and its velocity
gradient. 
At the region close to the wall of the inner cylinder,
on the other hand,
$s_1$ of case (i) is larger than those of the higher temperature cases (ii) and (iii), 
even though the velocity and velocity gradient of case (i) is significantly smaller
than those of the other two cases (see Fig. \ref{fig:Fig04}(a)).
The behaviors of $s_2$ for these three cases are opposite those of $s_1$.
From the behavior of $s_3$,
we can see that the chains are less-oriented along the $\phi$-direction
and relatively gentle compared with those along the $r$-direction
because the $\phi$-direction is the neutral direction.
In addition, the correlations 
between the deformation along the $\phi$-direction and those along the
other two directions, {\it i.e.,}
$S_{r \phi}^{\rR}$  and $S_{\phi z}^{\rR}$, are very small.
We can see that the deformation correlation
$S_{zr}^{\rR}$, on the other hand, is large.
These results are in good agreement with the polymer conformation images 
shown in Fig. \ref{fig:Fig06:160} and Fig. \ref{fig:Fig07:200}.
It is clearly understood that 
the chain deformation at a position 
is determined not only by the local velocity gradient
$\partial v_z/\partial r$
but also by the local-temperature-dependent relaxation time $\tau_{\rm d}(T(r))$,
{\it i.e.,} by the temperature dependent Weissenberg number
$\Wi_\rd(r,T)$ defined by eq.\eqref{eqn:WeissenbergNumber}. 

 \subsection{
  Dynamics after cessation of the external force density ($F_{\rm ext}=0$)
\label{sec:after_cessation}
}

\begin{figure}[t]
 \centering
\includegraphics[angle=-90,width=84mm]{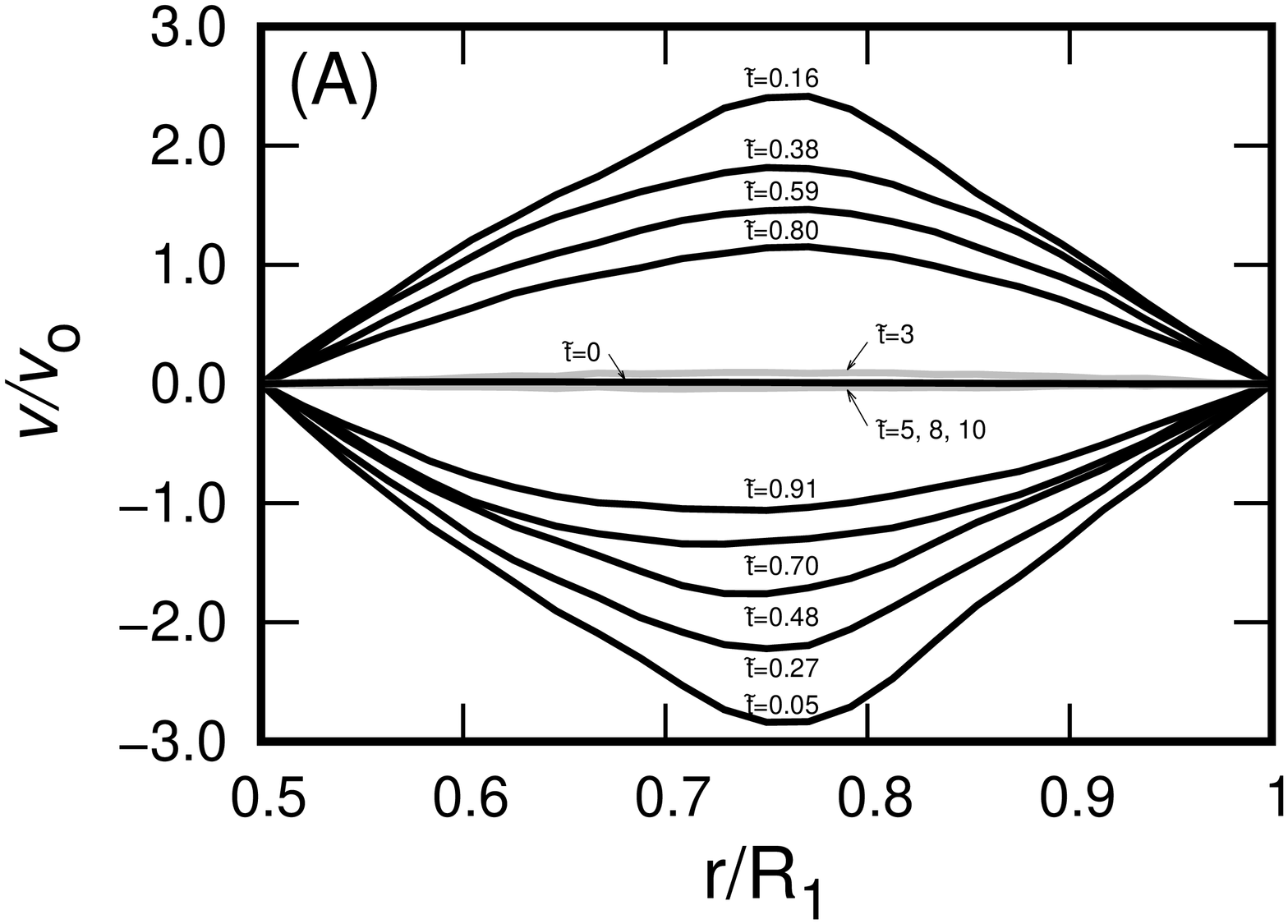}
\includegraphics[angle=-90,width=82mm]{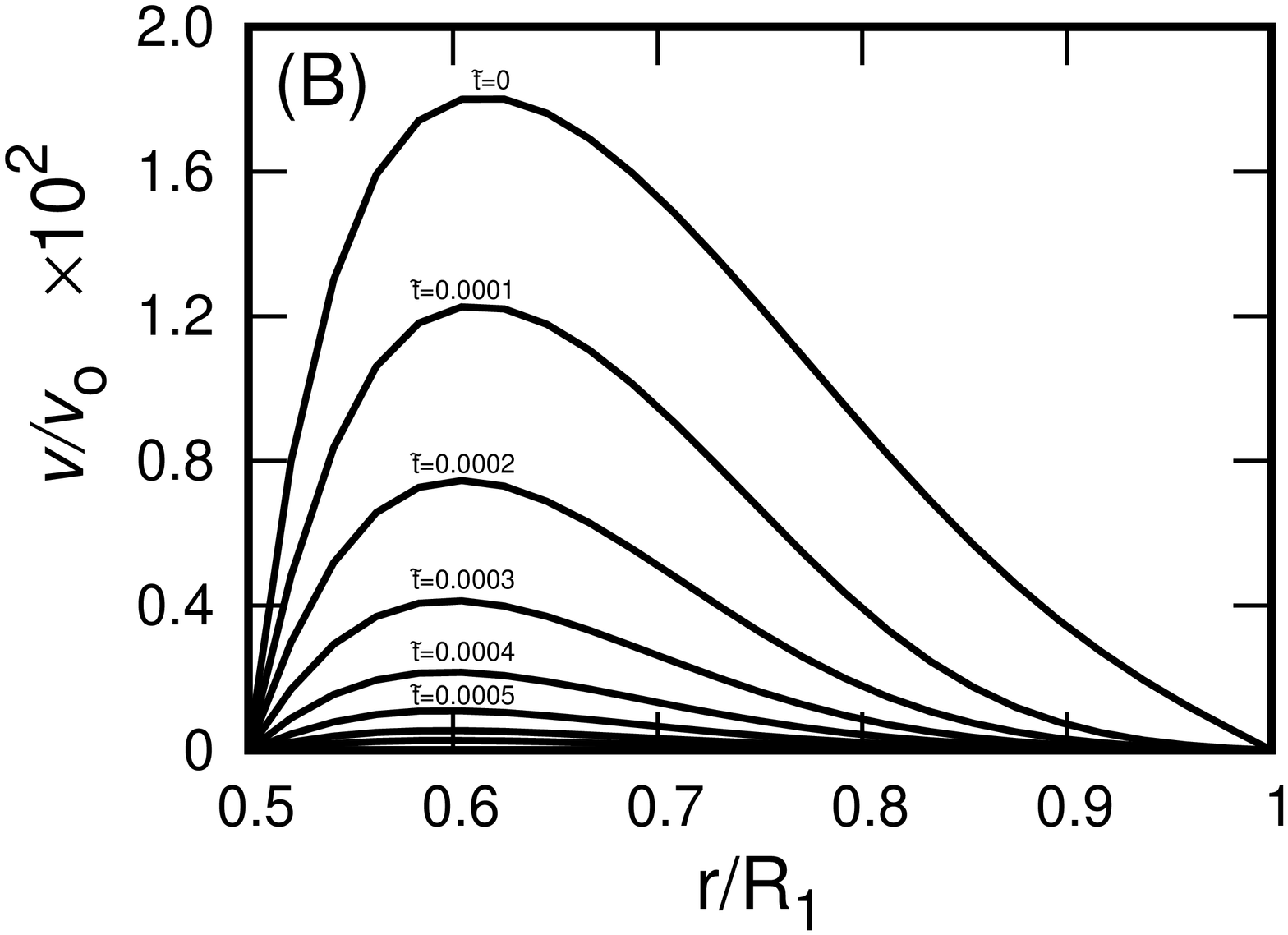}
\caption{
 Time evolutions of the velocity
 after cessation of the external force,
 by setting $F_{\rm ext}=0$
in  (A) the MSS system for case (iii) 200$^\circ$C in
 Fig.\ref{fig:Fig04},  
and (B) the corresponding non-Newtonian fluid (Cross-WLF model)
 evaluated in Fig.\ref{fig:Fig04}.
}
\label{fig:Fig10}
\end{figure}

\begin{figure}[t]
 \centering
\includegraphics[angle=-90,width=82mm]{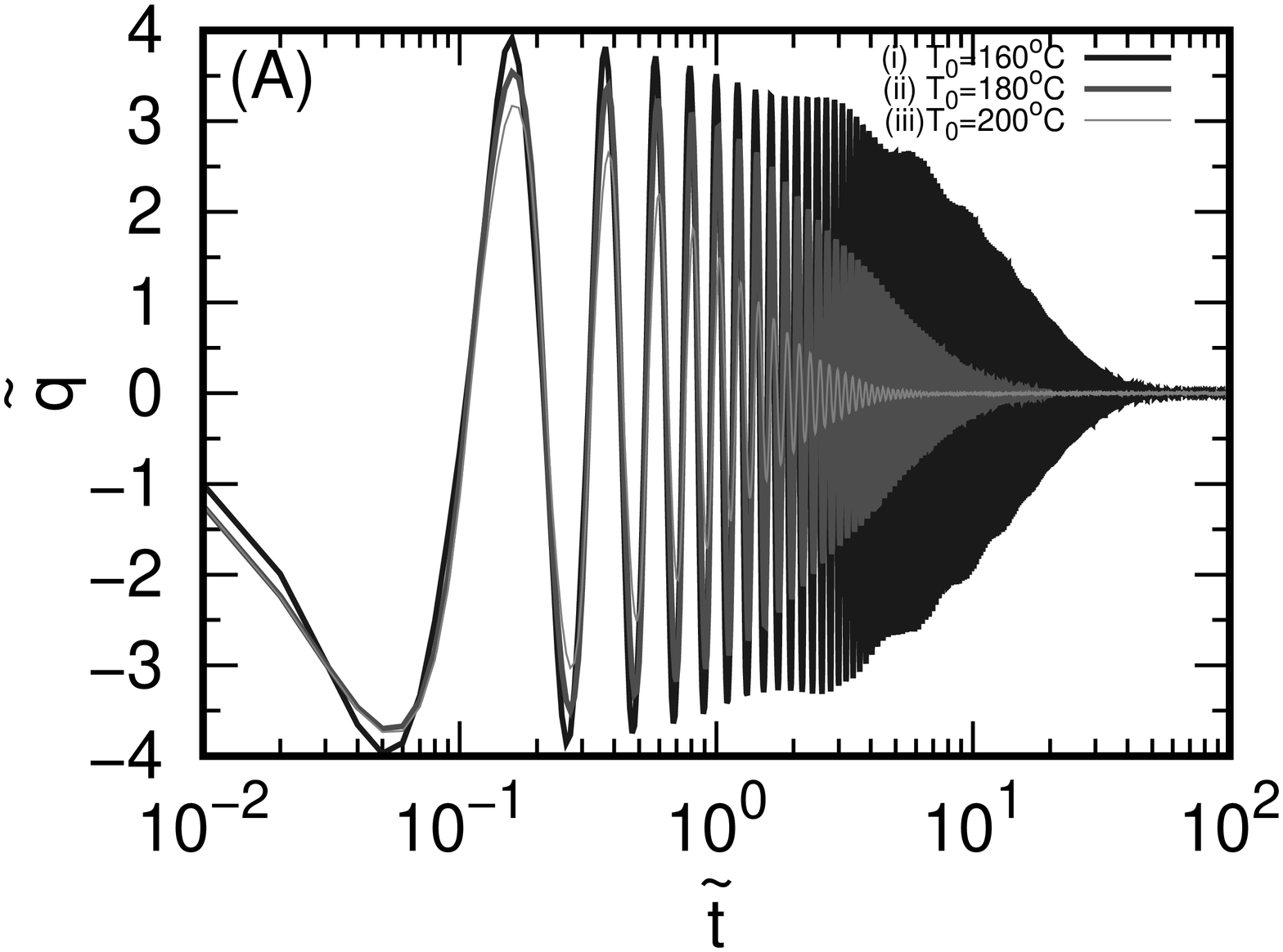}
\includegraphics[angle=-90,width=82mm]{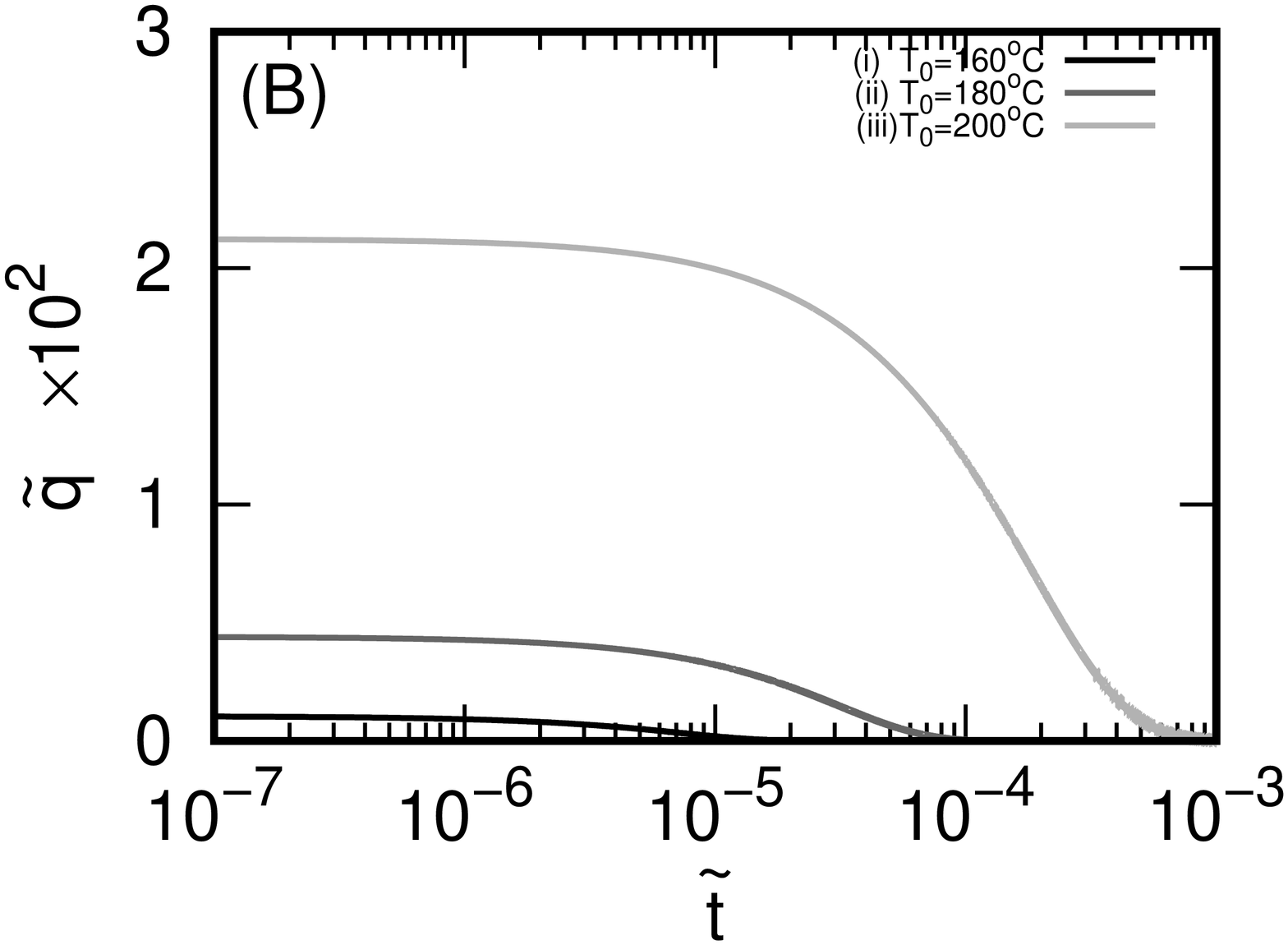}
\caption{
Scaled volumetric flow rates $\tilde q$
of  (A) MSS and (B) the corresponding non-Newtonian fluid (Cross-WLF model)
as a function of time 
for three different $T_0$ temperatures
(i)   $T_0=160^\circ$C, 
(ii)  $T_0=180^\circ$C, and 
(iii) $T_0=200^\circ$C.
 }
\label{fig:Fig11}
\end{figure}

In the previous subsection, we investigated the steady state of the
polymer melt flow between two coaxial cylinders under nonisothermal
conditions.
Here we focus on the dynamics of the polymer melt flow and its constituent
polymer chains after cessation of the external force density.
As the initial states for this investigation, we use
the steady states (macroscopic flow and temperature distributions, and
the microscopic state
of the polymer chains in each fluid element) obtained for the three
cases considered in the previous subsection,
(i)   $T_0=160^\circ$C, 
(ii)  $T_0=180^\circ$C, and 
(iii) $T_0=200^\circ$C, 
with the time set to $\tilde t=0$ when the force density is ceased.
The equations for the dynamics of the fluid velocity and temperature
are also given by \eqref{Eq_motion_nondim} and \eqref{Eq_energy_nondim},
respectively, but the apparent Reynolds number $\Re^{\rm (a)}$ and 
thermal P\'eclet number $\Pe^{\rm (a)}$ come into play in the dynamics,
in addition to $\Br^{\rm (a)}$.
As already mentioned, these three dimensionless parameters are evaluated to be
$\Re^{\rm (a)}=1.1 \times 10^{-2}$, 
$\Pe^{\rm (a)}=1.1 \times 10^5$, and
$\Br^{\rm (a)}=4.2 \times 10^4$.
The boundary conditions for the velocity and temperature are the same as
those for the steady state, {\it i.e.}, are also given by
\eqref{eqn:BC_for_v} and \eqref{eqn:BC_for_T}, respectively.
The treatment of the microscopic dynamics of the polymer chains is 
the same as that described in Sec.~\ref{subsec:micro} and Appendix I. 
\\
\indent
Fig.\ref{fig:Fig10} shows the time evolution of the flow for the case 
(iii) $T_0=200^\circ$C 
after cessation of the external force density for two types of
simulations: (A) the MSS for the polymer melt
and          (B) the simulation of a pure viscous fluid whose viscosity
is given by eq.\eqref{eqn:Cross_WLF_model}.
Note that the velocity and temperature distributions at the steady state
of (B) are the same as those of (A), 
as explained in Sec.~\ref{sec:steady state}.
 Fig.\ref{fig:Fig11} shows the scaled volumetric flow rates $\tilde q$
as a function of time for all three cases (i)-(iii).
This scaled volumetric flow rate $\tilde q = q /(v_0 R_1^2) $
is defined by the following equation as:
\begin{equation}
\tilde q = 2 \pi \int_{\tilde R_0}^{\tilde R_1} \tilde v_z(\tilde r) \tilde r d \tilde r.
 \label{eqn:flow_rate}
 \end{equation}
%
%
\begin{figure*}[t]
 \begin{center}
\includegraphics[angle=0,width=160mm]{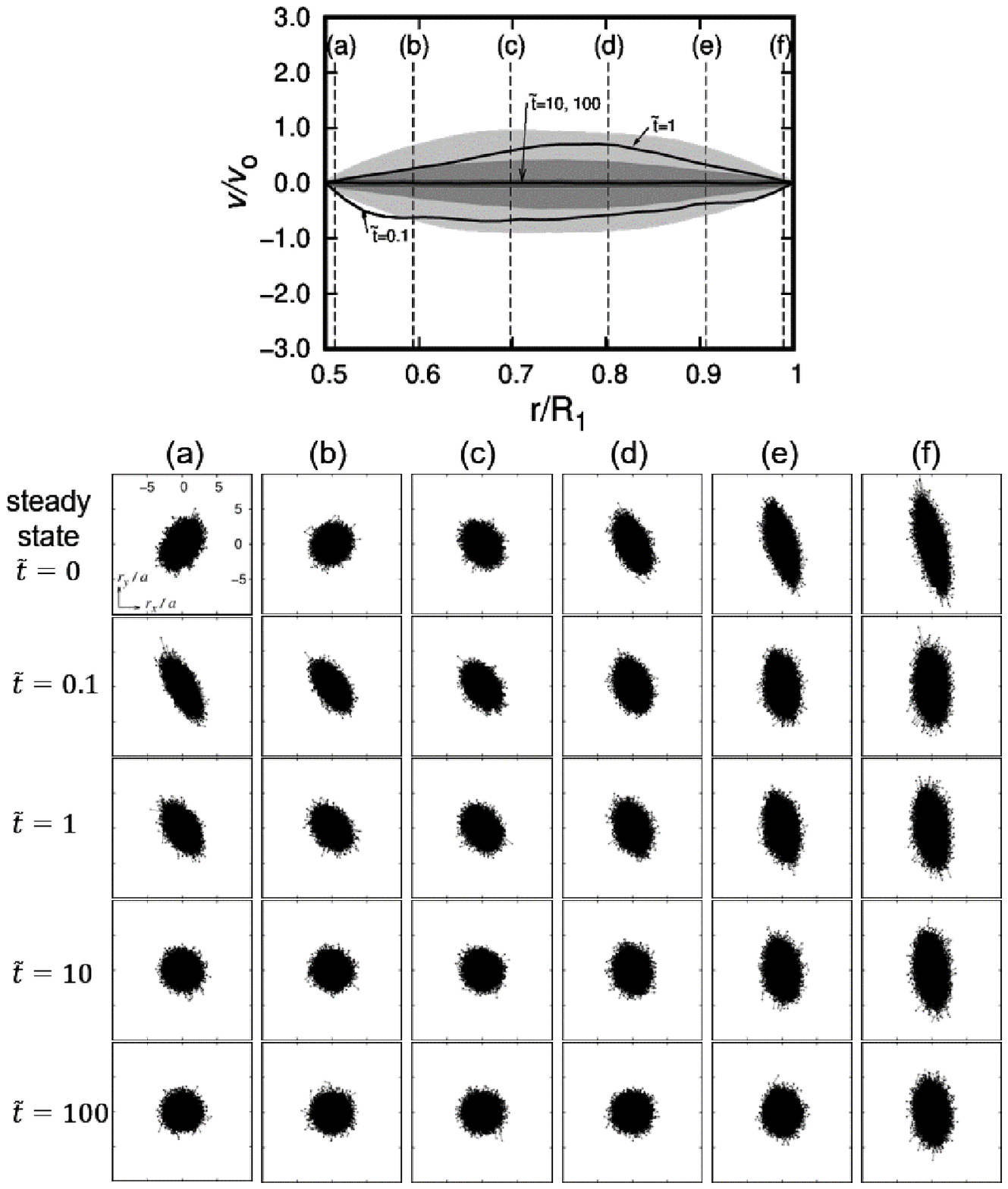}
 \end{center}
\caption{
 Velocity profiles for case (iii) at $\tilde t=0.1, 1, 10$, and  $100$ in the
 top panel. The light, intermediate, and dark gray regions are 
 drawn as the superposition of lines at 
 $1 < \tilde t \le 2$, $2 < \tilde t \le 5$, and
 $5 < \tilde t \le 10$, respectively.
 In the bottom panel, 
 local chain conformations 
 at $\tilde t=0.1, 1, 10$, and $100$ in the fluid element 
 located at the positions (a)-(f) in the top  panel are shown.
 }
\label{fig:Fig12}
\end{figure*}
We can see from Fig.\ref{fig:Fig10} and Fig.\ref{fig:Fig11} that the
polymer melt systems (A) clearly exhibit 
an oscillatory flow behavior with a decaying amplitude, on the other hand, 
the pure viscous system (B) only shows a flow-decaying-behavior and
reaches to a quiescent state in a short time.
The flow behavior in (A) can be characterized by three quantities, 
the amplitude decay-time $\d \tilde t_{\rm decay}^{\rm (viscoelastic)}$, 
the initial maximum amplitude $\cal A$, and
the period of oscillation ${\cal T}_{\rm osc}$,
whereas 
the flow behavior in (B) can be characterized solely by
the decay-time $\d \tilde t_{\rm decay}^{\rm (viscous)}$.
From Fig.\ref{fig:Fig10}(A) and Fig.\ref{fig:Fig11}(A),
we can see that the 
$\d \tilde t_{\rm decay}^{\rm (viscoelastic)}$
for
(i)   $T_0=160^\circ$C, 
(ii)  $T_0=180^\circ$C, and 
(iii) $T_0=200^\circ$C
are approximately
(i) 50, (ii) 20, and (iii) 5, respectively.
From Fig.\ref{fig:Fig10}(B) and Fig.\ref{fig:Fig11}(B)
we can see that 
$ \d \tilde t_{\rm decay}^{\rm (viscous)} \sim $
(i)   $ 10^{-5}$,  
(ii)  $ 10^{-4}$
and
(iii) $ 10^{-3}$,
and the decay-time becomes longer as $T_0$ increases, 
which is opposite to the $T_0$-dependence of the decay-time in (A).
As explained in Appendix II,  
the short decay time in the pure viscous system (B)
can be estimated by 
$\d \tilde t_{\rm decay}^{\rm (viscous)}
 \sim \Re^{\rm (a)}/[(2\pi)^2 \tilde \eta_0(\bar T)]$, 
where $\bar T$ is an effective temperature.
For example, if $\bar T$ is approximated 
to be the temperature of the inner wall,
{\it i.e.,} $\bar T \sim T_0$,
the $\d \tilde t_{\rm decay}^{\rm (viscous)}$
are roughly estimated as 
 (i)  $ 3.5  \times 10^{-6}$,  
(ii)  $ 0.97 \times 10^{-4}$,
and 
(iii) $ 1.0\times10^{-3}$
by using
$\tilde \eta_0(160^\circ$C$)\simeq 79$,
$\tilde \eta_0(180^\circ$C$)\simeq 2.8$,
$\tilde \eta_0(200^\circ$C$)\simeq 0.27$,
and $\Re^{\rm (a)} \sim 1.1 \times 10^{-2}$,
and these estimated values are consistent
with the results of the simulations mentioned above.
Regarding the amplitude decay-time in the polymeric system, 
we can understand that the decay-time
is approximately twice the polymer relaxation time, {\it i.e.},
$2\tilde \tau_\rd(T) \sim 2 a_T \tilde \tau_\rd(T_{\rm ref})$
as shown in Appendix II. 
So we can estimate $\d \tilde t_{\rm decay}^{\rm (viscoelastic)}$ as 
(i)   $2\tilde \tau_\rd(160^\circ$C$)\simeq 732$,
(ii)  $2\tilde \tau_\rd(180^\circ$C$)\simeq 25.6$ and
(iii) $2\tilde \tau_\rd(200^\circ$C$)\simeq 2.4$, 
which roughly grasps the tendency of the decay-times in (A).
We can also understand the opposite tendency of the decay-time
as a function of $T_0$ between (A) and (B)
through the temperature dependence of $\eta_0(T)$ and $\tau_\rd(T)$.
It should be noted that the temperature distributions in both cases (A) and (B)
show almost no change on
the time scale of the flow decay
($\tilde t \lesssim \d \tilde t_{\rm decay}^{\rm (viscoelastic)}$), 
with respect to the steady state temperature distribution shown
in Fig.\ref{fig:Fig04}(b), 
because of the very large thermal P\'eclet number $\Pe^{\rm (a)}$.
The characteristic temperature relaxation time
to the equilibrium distribution given in eq.\eqref{eqn:analytic_solution_of_T}
is estimated to be $\d \tilde t_{\rm relax}^{(T)} \sim \Pe^{\rm (a)} \simeq 10^5$.
The period of the oscillating flow $\tilde {\cal T}_{\rm osc}$ for
case (iii) is evaluated visually to be
$\tilde {\cal T}_{\rm osc} \simeq 0.2$  from Fig.\ref{fig:Fig10},
and it can be found that those for the other two cases (i) and (ii)
are almost the same, as seen from Fig.\ref{fig:Fig11}(A).
As given in Appendix II, 
the period is theoretically estimated to be
$\tilde {\cal T}_{\rm osc} \simeq 2\pi/\tilde \omega_1$
where 
$\tilde \omega_1 \equiv
2\pi [\tilde \eta_0(\bar T)/
\Re^{(\rm a)}
\tilde\tau_\rd(\bar T)
]^{1/2}$.
By using this expression,
the periods $\tilde {\cal T}_{\rm osc}$
for (i)-(iii) are evaluated 
to be (i) 0.23, (ii) 0.22, (iii) 0.21. 
Interestingly, the amplitude of the oscillating flow 
is approximately one hundred times larger than 
the magnitude of the flow at the steady-state.
The large amplitude of the oscillation comes from the releasing of 
the stored elastic energy in the deformed polymer chains.
The maximum amplitude ${\cal A}$ in the first oscillation
after the cessation of $F_{\rm ext}$ is larger for smaller $T_0$, 
it can estimated theoretically as
${\cal A} \sim 4 \tilde F_{\rm ext} /\left( \pi \tilde \omega_1(\bar T)
\Re^{(\rm a)}\right)$,
and is evaluated to be
(i)    2.71
(ii)   2.65
and 
(iii)  2.59, 
which is consistent with the results shown in Fig.\ref{fig:Fig10}(A).
This tendency can be attributed to the longer relaxation times
for stretch and orientation of polymer chains at lower temperatures.
\\ \indent
To understand the relation between the oscillatory flow and 
the chain deformation, 
the velocity profiles at
$\tilde t=0.1$, $1$, $10$, and $100$ for case (iii) 
are plotted in the top panel of Fig.\ref{fig:Fig12}.
In addition,
velocity profiles at $1 < \tilde t \le 2$, $2 < \tilde t \le 5$, and 
$5 < \tilde t \le 10$ are superimposed, which results in 
the light, intermediate and dark gray regions, respectively.
In the bottom panes of Fig.\ref{fig:Fig12}, 
local chain conformations 
at the positions corresponding to (a)-(f) in the top panel, 
and also in Fig.\ref{fig:Fig07:200},
at $\tilde t=0,~~0.1,~~1,~10$, and $100$ are 
drawn by fixing the center of mass of each chain to the origin.
At steady state ($\tilde t=0$),
polymer chains close to the inner wall, ({\it i.e.,}(a)),
are stretched along the flow direction and slightly tilted in
the positive radial direction, 
located near the center region ((b) and (c))
are almost isotropic, 
and those close to the outer wall ((d), (e) and (f)) are highly stretched
and tilted toward the inner side,  
because of the long relaxation time
coming from the lower temperature of the outer wall.
After the cessation of the external force, at first, 
elastic energy stored in the chains induces
a backflow, as seen in Figs.\ref{fig:Fig10}(A) and \ref{fig:Fig11}(A), 
with the chains oriented in the backflow direction at $\tilde t \lesssim 0.05$
(although we did not show the corresponding figures in Fig.\ref{fig:Fig12}).
At around $\tilde t \simeq 0.05$, 
the backflow starts decreasing and 
reverting to a flow in the positive direction.
At $\tilde t = 0.1$ it is still on its way back towards a positive flow,
with the chains still oriented in the backflow direction, as can be seen
in the second row of the chain conformation panels. 
At $\tilde t \simeq 0.16$,  the velocity reaches a maximum positive flow.
Then, the flow again exhibits a backflow
and reaches the second largest backflow at $\tilde t=0.27$.
After that, the oscillatory back- and forth- flow continue, 
with decreasing amplitude.
Polymer chains located at almost all positions
go back to the relaxed state
at around a time between
$\tilde t=1$ and $\tilde t=10$,
except for the chains located close to the outer wall, 
because of the longer relaxation time ($\tilde \tau_\rd(T_1)\simeq 366$)
at low temperatures ($T_1 \simeq 160^\circ$C).
Although the chain conformations close to the outer wall
are still oriented, to a certain extent, the macroscopic flow
has already decayed, as seen from Figs.\ref{fig:Fig10}(A),
\ref{fig:Fig11}(A) and 
\ref{fig:Fig12} (top panel).

\section{Conclusions\label{sec:Conclusion}}

We investigated a well-entangled polymer melt flow 
under nonisothermal conditions 
by using a multiscale simulation method.
Thus far, a multiscale simulation has been applied 
to isothermal flow problems; 
here, we extended the multiscale simulation method
such that it is applicable to the nonisothermal flow problem. 
At the macroscopic level, the way to incorporate the effect of temperature
into the macroscopic equations (momentum and energy balance equations)
has already been well established, and the method can be directly applied to flow problems
if the temperature-dependence of the density $\rho$,
the specific heat capacity $C_p$, and the thermal conductivity $k$
and the stress tensor field $\bm \sigma$ are known,
but at the microscopic level, the way to incorporate the effect 
into a coarse-grained microscopic model has not yet been clarified
because the temperature dependence of the parameters
used in this coarse-grained microscopic model is usually not known.
It is well known that temperature effects can be incorporated into
constitutive equations by scaling its parameters, based on
the time-temperature superposition rule. Given this fact,
we have used the same time-temperature superposition,
in reverse, in order to extend the
multiscale simulation method to nonisothermal flows.
In this way, we are able to evaluate the
temperature-dependent relaxation times and stresses needed for the
slip-link simulations at the microscopic level.

To demonstrate the effectiveness of the method, 
we investigated macroscopic polymer melt flows and
microscopic state polymer chains between two coaxial cylinders
under nonisothermal conditions, 
in which we used a polymer melt composed
of linear monodisperse polymer chains
with $Z$-entanglement points ($Z=Z_0+1$, $Z_0$ is the number of strands, and $Z_0=10$)
on a chain in equilibrium.
The temperatures of the inner and the outer cylinders are respectively set to
$T_0$ and $T_1$, and $T_1$ is always set to $T_1=160^\circ$C),
while $T_0$ is set to
        (i)   160$^\circ$C,
        (ii)  180$^\circ$C,
   or   (iii) 200$^\circ$C.
As a result, 
the extended method succeeds in 
simulating temperature-dependent flow behaviors
and creating a relation between 
the macroscopic flow and temperature distribution in the tube and
the microscopic states of the polymer chains along the radial direction. 
Although the polymer chains are stretched and oriented 
along the flow direction according the local shear rate, 
the degree of deformations of the polymer chains
are determined not only by the local shear rate
but also by the local temperature through the temperature-dependent
relaxation times $\tau_\rd$ and $\tau_\rR$.

At steady states under a constant external force, near the outer cylinder,
the length of primitive path $L$
and the number of entanglements on a single chain $Z$
in the high-temperature cases (ii) and (iii)
become longer and smaller than those in case (i), respectively.
Near the inner cylinder,
on the other hand,
these quantities display opposite tendencies,
although the magnitude of the shear rates in this region
is larger than those in the region near the outer cylinder.
These behaviors can be explained by considering 
the temperature-dependent reptation-time-based Weissenberg number
$\Wi_\rd(r,T)$, and therefore, the number 
is found to be a very effective parameter
in understanding the microscopic state of polymer chains.

We also investigated the dynamics of the system after cessation of the
external force, by using the steady state as an initial state.
The flows of the polymeric system exhibit an oscillatory decaying flow,
whereas the corresponding pure viscous system
exhibits a simple decaying flow.
We also demonstrated how the polymer conformations
are related to the oscillatory decaying flow behavior.

In this paper, although we could not directly compare our numerical results with
experimental ones,
we are hoping to make intensive comparisons of the data obtained by
the nonisothermal MSS and experimental data
in the future, 
not only from the viewpoint of the macroscopic properties,
such as the flow and temperature distributions,
but also from the microscopic state of the polymer chains. 
We will continue to apply the present method to more industrially oriented problems 
to predict the temperature-dependent flow behavior
and microscopic state of a polymer chain.
Finally, we believe that the present method offers us a new idea 
and/or novel knowledge not only on macroscopic flow but also a
microscopic state of polymer chains under nonisothermal conditions.

\section*{
 Acknowledgments
 }

This work was supported partially by JSPS KAKENHI Grant Number 19H01862,
the Ogasawara Foundation for the Promotion of Science and Engineering 
and MEXT as the ``Exploratory Challenge on Post-K computer''
(Frontiers of Basic Science: Challenging the Limits).

\vspace{0.5cm}
\noindent
{\bf
Appendix I:\\
Temperature dependent update of the slip-link primitive path
}

We explain the procedure for updating the state of a primitive path
and slip-links of a polymer in a fluid particle 
at a temperature $T$, as shown in Fig.\ref{fig:Fig03}.
The same procedures mentioned below in (i)-(v) are applied
to all the primitive paths and slip-links 
of $N_\rp$-polymer chains in the fluid particle simultaneously.
\begin{enumerate}
\item[(i)] Obtain the temperature $T(t)$ and velocity gradient tensor
           $\bkappa(t)$ evaluated macroscopically at the position
           of the fluid particle and at a time $t$.

\item[(ii)] Apply an affine transformation to the positions
            of slip-links according to the velocity 
            gradient tensor $\bkappa (t)$ obtained at item (i).

\item[(iii)] Update the length of the primitive path
             after a time duration $\Dt$
             from the time $t$ according to the following equation:
             \begin{eqnarray}
              && L(t+\D t) = L(t)  
                    - \frac{1}{\tau_{\rm R}(T)}
                      \left(L \left(t \right) - L_0 \right) \D t
                      \nonumber \\ 
             && \qquad\qquad\qquad\qquad
                    + \D L_{\rm affine}
                    + \sqrt{2 a^2 \D t \over 3 Z_0 \tau_\re(T)} w_\rL,
             \label{eqn:L}      
             \end{eqnarray}
             where  
             $\tau_{\rm R}(T)(= Z_0^{2}\tau_{\rm e}(T))$ and $\tau_{\rm e}(T)$
             are the Rouse relaxation time of a polymer with length $L_0$
             and that of a polymer with length $a$ at $T$, respectively.  
             $\tau_{\rm e}(T)$ is expressed by $\tau_{\rm e}(T_{\rm ref})$ at
             a reference temperature $T_{\rm ref}$ through the following
             relation:
             \begin{equation}
               \tau_{\rm R}(T)=a_T \tau_{\rm R}(T_{\rm ref}). 
               \label{tau_R}
             \end{equation} 
             Here, $a_T$ is the shift factor 
             that appears in the TTS rule
             and is given in eq.\eqref{eqn:WLF}.
             $\D L_{\rm affine}$ is the contribution from an affine
             deformation, and $w_\rL$ is a Gaussian random number
             with zero mean and unit variance. 
\item[(iv)] Update the lengths of the two free ends ($s_\pm$) by a reptation motion.
            The lengths $s_\pm$ are updated as 
            \begin{eqnarray}
              \D s_{\pm} &=& s_\pm(t+\D t) - s_\pm(t) \nonumber \\
                         &=& {1 \over 2} [ \D L - \D L_{\rm affine}] 
                            \pm \sqrt{2D_\rc \D t~} w_\rs
             \label{eqn:Reptation}
            \end{eqnarray}
            where $w_\rs$ is a Gaussian random number with zero mean
            and unit variance, 
            $D_\rc=L_0^2 /\pi^2 \tau_\rd^{\rm (DE)}(T) $.

            The first term on the right-hand side of eq.\eqref{eqn:Reptation}
            is the contribution from the change in the contour length
            except for the affine deformation,  
            and the second term originates from the reptation motion,
            where
            $
            \tau_\rd^{\rm (DE)}(T)=a_T
            \tau_\rd^{\rm (DE)}(T_\rmref)
            $
            and
            $
            \tau_\rd^{\rm (DE)}(T_\rmref)=3Z_0^3 \tau_\re(T_\rmref)$\cite{Doi1986}.
\item[(v)] Creation and removal of slip-link(s).  
           The number of entanglements $Z$ on a chain is changed
           by the motions of its chain ends and
           the motions of other chains entangled
           with this chain (constraint release; CR).
           When the length of a chain end becomes less than zero,
           the slip-link on the chain end is removed.
           At the same time, 
           the partner slip-link to the removed one is also removed (CR).
           Conversely, when the length of a chain end becomes larger than $a$,
           a new slip-link is created on it.
           Simultaneously, a polymer chain is randomly selected from
           other polymer chains.
           A strand on the chain to be entangled is chosen with
           a probability proportional to the length of the strand, 
           and a new slip-link is created on the strand. 
\end{enumerate}
After updating the states of the polymer chains, 
the stress tensor at the position of the fluid particle
can be evaluated by:
\begin{equation}
\sigma_{\alpha\beta}
 ={b_T} \sigma_{\rm e}(T_{\rm ref})
 \left\langle
                       \frac{r_{\alpha}r_{\beta}}
                       {a|{\br} |}\right\rangle,
\quad \alpha,\beta \in \{x, y, z \}
\label{evaluation_of_sigma}
\end{equation}
where $\br$ is a connector vector between two adjacent slip-links and 
$a$ and $\sigma_{\rm e}(T_{\rm ref})$ are units of the length and stress in the slip-link model,
respectively. 
The bracket $\langle (\cdots) \rangle$ means the statistical average of $(\cdots)$.
As seen from the bulk rheological data at different temperatures (see
Figs.~\ref{fig:Fig02}(a) and (b)),
the method in which the time-temperature superposition rule is inversely
used can successfully take into account the temperature effect on the stress.
As seen from eq.\eqref{eqn:WLF}, 
$a_T$ changes exponentially as a function of the temperature,
the dynamics of the primitive paths, {\it i.e.},
the time scales of the length change and the motions of the primitive paths,
are significantly altered
through the reptation time $\tau_\rd(T)$
and the Rouse relaxation time $\tau_\rR(T)$. 
Thereby, 
the change in temperature highly influences
the state of the primitive path and the resultant stress tensor.

\vspace{1cm}
\noindent
{\bf
Appendix II:\\
Approximated transient dynamics after
the cessation of the force density
}
\\
\indent
To understand the transient dynamics of a polymeric flow
from a steady state to a quiescent state,
after the cessation of the force density $F_{\rm ext}$, 
we simplify eq.\eqref{Eq_motion_nondim}
by replacing the temperature-dependent physical parameters
with effective constants and 
neglecting the curvature effect.
In addition, the stress is assumed to be given by a Maxwell
constitutive equation with the relaxation time $\tau_\rd(\bar T)$
and the viscosity $\eta_0(\bar T)$ at an effective temperature $\bar T$.
For simplicity, we express ${\tilde v}_z$ by $\tilde v$ 
and $\tilde{\sigma}_{rz}$ by $\tilde \sigma$, then 
eq.\eqref{Eq_motion_nondim} and
the equation for the stress are written as
\begin{eqnarray}
{\Re^{\rm (a)}} \frac{ \partial {\tilde v} (\tilde{r}, \tilde{t})}{\partial \tilde{t}}
&=&
     {\partial \tilde{\sigma} \over \partial \tilde{r} }
   + \left ( 1 - \Theta(\tilde t) \right ){\tilde{F}_z}^{\rm {(ext)}}
\label{Eq_motion_nondim_approximated}
\\
\tilde \tau_\rd(\bar T)
{ \partial {\tilde \sigma}(\tilde{r}, \tilde{t})
   \over \partial \tilde{t} }
&=&
- {\tilde \sigma}(\tilde{r}, \tilde{t})
+ \tilde \eta_0(\bar T) {\partial {\tilde v} \over \partial \tilde{r} }.
   \label{Stress_approximated}
\end{eqnarray}
where $\Theta(\tilde t)$ is unity for $\tilde t \ge 0$ and zero for
$\tilde t < 0$.
The initial conditions ($\tilde t=0$)
for $\tilde v$ and $\tilde \sigma$
are given by the steady state of
eqs.\eqref{Eq_motion_nondim_approximated} and \eqref{Stress_approximated}, 
{\it i.e.}, $\tilde v(\tilde r,\tilde t=0)=\tilde v_{\rm s}(\tilde r)$
and $\tilde \sigma=\tilde \sigma_{\rm s}(\tilde r)$,
where
$\tilde v_{\rm s}(\tilde r)
=\tilde F_{\rm ext}(\tilde r -1/2)(1- \tilde r)/2\tilde \eta_0(\bar T)$
and
$\tilde \sigma_{\rm s}(\tilde r)=\tilde F_{\rm ext}(-\tilde r+3/4)$.
The solution of $\tilde v$
of eqs.\eqref{Eq_motion_nondim_approximated} and \eqref{Stress_approximated}
is 
\begin{eqnarray}
&& \tilde v(\tilde r,\tilde t)  =
      \exp \left  (
                    - {\tilde t  \over 2 \tilde \tau_\rd } 
          \right )
 \sum_{n=1,3,5\cdots}^\infty 
    { \sin \Bigr ( \pi n \left (2\tilde r - 1 \right ) \Bigr ) \over
    \pi^3 n^3 }
\nonumber
\\ 
&& \qquad
     \times
     \biggr [
                           \cos \left ( \tilde \omega_n \tilde t \right )
            +          S_n \sin \left ( \tilde \omega_n \tilde t \right )
     \biggr ] \quad \text{for}~ \tilde t \ge 0
\\
&&
\tilde \omega_n
      = \left [
          { 4 \pi^2 n^2 \over \Re^{(\rm a)} }
          { \tilde \eta_0 \over \tilde \tau_\rd }
        - {1 \over (2\tilde \tau_\rd)^2}
        \right ]^{1/2}
        \simeq
          2 \pi n
          \Bigr ( { \tilde \eta_0 \over
          \tilde \tau_\rd \Re^{(\rm a)}  }
          \Bigr )^{1/2}, \qquad
\\          
&& S_n =  
           \left[
            - 
            {4 \pi^2 n^2 \over \Re^{(\rm a)}}
            \tilde F_{\rm ext}
            + {1 \over 2 \tilde \tau_\rd}
          \right]{ 1 \over  \tilde \omega_n }
       \simeq
        - {4 \pi^2 n^2 \tilde F_{\rm ext} \over
          \tilde \omega_n \Re^{(\rm a)}  } .
\quad
\end{eqnarray}
From the above expressions,
$\tilde \eta_0 / \Re^{(\rm a)},~
 {\tilde F_{\rm ext} / \Re^{(\rm a)}} 
\gg
1/\tilde \tau_\rd$ and the fact that $n=1$ is the  dominant mode, 
the amplitude decay-time $\d \tilde t_{\rm decay}^{\rm (viscoelastic)}$,
the maximum amplitude in the first oscillation ${\cal A}$,
and 
the period of the oscillating flow $\tilde {\cal T}_{\rm osc}$
are estimated as
\begin{eqnarray}
&& \d \tilde t_{\rm decay}^{\rm (viscoelastic)} \simeq 2 \tau_\rd(\bar T),
\\
&&
 {\cal A} \simeq {4 \tilde F_{\rm ext} \over
\pi 
\tilde \omega_1
\Re^{(\rm a)}
},
\\
&&
\tilde {\cal T}_{\rm osc} \simeq {2\pi \over \tilde \omega_1 }.
 \end{eqnarray}
\indent
The equation for the pure viscous fluid
with viscosity $\tilde \eta_0(\bar T)$
is given by putting $\tilde \tau_\rd=0$ in eq.\eqref{Stress_approximated}.
The analytic solution for the viscous fluid is given as
\begin{eqnarray}
&& \tilde v(\tilde r,\tilde t)  = \sum_{n=1,3,5\cdots}^\infty 
 { 2 \tilde F_{\rm ext} \over \tilde \eta_0 }    
    { \sin \Bigr ( \pi n \left (2\tilde r - 1 \right ) \Bigr ) \over
    \pi^3 n^3 }
\nonumber    
\\ 
&& \qquad \qquad 
     \times
     \exp \left  (
                    - 4 \pi^2 n^2 {\tilde \eta_0 \over \Re^{\rm (a)}} \tilde t~
          \right ).
\label{eqn:solution_for_pure_viscous_fluid}          
\end{eqnarray}
From eq.\eqref{eqn:solution_for_pure_viscous_fluid},
the decay time in the pure viscous system (B)
can be estimated as
\begin{equation}
\d \tilde t_{\rm decay}^{\rm (viscous)}
 \sim {\Re^{\rm (a)} \over (2\pi)^2 \tilde \eta_0}.
\end{equation}



\end{document}